%% file: BI-SMB.tex
\documentclass[authoryear,10pt]{article}

\usepackage{natbib}
\usepackage{fullpage}
\usepackage{authblk}
\input{preamble.tex}

\graphicspath{ {images/} }
\begin{document}

    \title{Model-based process design of a ternary protein separation using multi-step gradient ion-exchange SMB chromatography}

    \author[1,2]{Qiao-Le He}
    \author[1,2]{Liming Zhao}
    \affil[1]{\textit{State Key Laboratory of Bioreactor Engineering, East China University of Science and Technology, 200237 Shanghai, China}}
    \affil[2]{\textit{R\&D Center of Separation and Extraction Technology in Fermentation Industry, East China University of Science and Technology, 200237 Shanghai, China}}
    \date{Apr.~15, 2020}

    \maketitle

    \begin{abstract}
        Prominent features of simulated moving bed (\textsc{smb}) chromatography processes in the downstream processing is based on the determination of operating conditions.
        However, effects of different types of uncertainties have to be studied and analysed whenever the triangle theory or numerical optimization approaches are applied.
        In this study, a Bayesian inference based method is introduced to consider the uncertainty of operating conditions on the performance assessment, of a glucose-fructose \textsc{smb} unit under linear condition.
        A multiple chain Markov Chain Monte Carlo (\textsc{mcmc}) algorithm (\ie, Metropolis algorithm with delayed rejection and adjusted Metropolis) is applied to generate samples.
        The proposed method renders versatile information by constructing from the \textsc{mcmc} samples, \eg, posterior distributions, uncertainties, credible intervals of the operating conditions, and posterior predictive check, and Pareto fronts between each pair of the performance indicators.
        Additionally, the \textsc{mcmc} samples can be mapped onto the $(m_\text{II}, m_\text{III})$ and $(m_\text{IV}, m_\text{I})$ planes to show the actually complete separation region under uncertainties.
        The proposed method is a convenient tool to find both optimal values and uncertainties of the operating conditions.
        Moreover, it is not limited to \textsc{smb} processes under the linear isotherm; and it should be more powerful in the nonlinear scenarios.

    \end{abstract}


\section{Introduction}
Purification and separation are major concerns in the downstream processing of industries \citep{carta2010protein,scopes2013protein}.
Chromatography is a prevailing purification and separation technology \citep{guiochon2002preparative}.
Simulated moving bed (\textsc{smb}) \citep{broughton1961continuous} as a continuous chromatographic separation technology is an alternative to the conventional batch chromatography, since \textsc{smb} processes have characteristic features (\eg, high productivity and low solvent consumption) \citep{seidel2008new,rajendran2009simulated}.
It has been widely applied to separations of fine petrol-chemicals, sugars, and pharmaceuticals \citep{juza2000simulated,rajendran2009simulated,faria2015instrumental}.
%

There are three types of issues in process designs of \textsc{smb} systems.
The first type deals with the determination of network configuration (\eg, total columns and optimum number of columns in individual zones), the geometry of columns (\eg, length, diameter, size distribution of beads).
The second type arises in measurement of experimental parameters (\eg, axial dispersion coefficient, porosities, dead volumes and adsorption parameters).
Optimization of operating conditions (\eg, zonal flowrates, switching time), with respect to studied objectives, appears in the third type.
In designs of \textsc{smb} separations, the key issue is the determination of the optimum values of operating parameters, that is, assuming that the network configuration, the column geometry and the packing parameters have been fixed, that the adsorption equilibrium properties of the components have been known from the column model calibration.

Classical algebraic methods to the selection of operating conditions are McCabe-Thiele diagram \citep{ruthven1989counter}, the safety margin approach \citep{mihlbachler2001effect}, the standing wave design \citep{ma1997standing,mallmann1998standing}, and the triangle theory \citep{storti1993robust,storti1995design}.
The safety margin approach is, in principle, equivalent to the triangle theory, which is derived based on the equilibrium theory model (\ie, ideal column model with the linear isotherm).
Initially, the triangle theory was barely suitable for determining zonal flowrates of \textsc{smb} processes under the linear isotherm condition.
It was later extended to processes described by Langmuir pattern nonlinear isotherms \citep{mazzotti1997optimal}.
Then, versatile results on triangle theory for the design of \textsc{smb} processes have been published \citep{nowak2012theoretical, kim2016combined, lim2004optimization, kazi2012optimization, bentley2013prediction, bentley2014experimental, sreedhar2014simulated, toumi2007efficient, silva2015modeling, kiwala2016center}.
No research has yet been reported to extend the triangle theory to nonlinear isotherms, such as spreading model \citep{ghosh2013zonal,ghosh2014zonal}, steric mass-action \citep{brooks1992steric}.
The McCabe-Thiele diagram can be applied to processes described by any kind of adsorption isotherms, but is limited to binary separations.
The standing wave design method that was originally suitable for linear and ideal \textsc{smb} processes \citep{ma1997standing} was extended to nonlinear and nonideal processes by considering the axial dispersion effects and mass transfer resistances with correlation terms to the design equations \citep{xie2000extended}.
Further, the effects of dead volumes and pressure drops that are vital in real applications were taken into account \citep{lee2005standing}.

Although the above algebraic solutions are powerful tools for determination of operating conditions of \textsc{smb} processes, more insights should be shed into the interpretation of rate-limiting mass transfer of the most current macro-molecules.
Moreover, substances separated by \textsc{smb} processes have recently been evolved from monosaccharides to macro proteins, which can undergo conformation, orientation changes and aggregation in the process.
The thermodynamics of macro-molecules, as observed both in experiments \citep{clark2007new} and molecular dynamics \citep{dismer2010structure, liang2012adsorption, lang2015comprehensive}, are much more complicated than that described by the linear isotherm, even by the Langmuir pattern or the steric mass-action  model.
Therefore, finding operating conditions that guarantee the required performance indicators under nonlinear and nonideal conditions is a major challenge in \textsc{smb} systems.

Numerical solutions can be an alternative to the theoretical algebraic solutions, to constitute the chromatograms of various network configurations, to calculate the performance indicators, to search for the optimum flowrates and switching time \citep{rodrigues2007experimental,agrawal2012comparison,nowak2012theoretical,wu2013improving,li2014using,li2014model,bentley2014experimental,sreedhar2014simulated,yao2015combination,yu2015simulation}.
Numerical solutions of different column models with various adsorption isotherms have been investigated \citep{dunnebier2000optimal}.
The characteristic features of \textsc{smb} processes can be further improved by systematically tuning the column configurations and operation modes \citep{zhang2002multiobjective}.
However, process designs of \textsc{smb} units in industries do not fully benefit from the model-based approaches \citep{briskot2019prediction}.
This is mainly attributed to the lack of uncertainty analysis in process designs of \textsc{smb} systems \citep{kurup2008monte,borg2013effects}.

In order to efficiently model, design, and control \textsc{smb} processes, any type of uncertainties (\eg, from equipment calibration, signal detectors, model assumption, model calibration, numerical solution) that could affect the performance assessment should be taken into account.
Without consideration or inaccurate estimation of dead volumes results in errors in the evaluated adsorption parameters \citep{samuelsson2008impactI,grosfils2010parametric} and in the selection of adsorption model by chromatogram fitting \citep{samuelsson2008impactII}.
The effects of concentration measurement error \citep{joshi2006model} and detector disturbance \citep{zhang2001numerical} on the estimation of adsorption parameters have been studied.
\citet{borg2013effects} analysed the quantitative effect of uncertainty in the experimental conditions (Monte Carlo sampled experimental conditions) on the parameter estimation of a chromatographic column model.
\citet{kurup2008monte} presented a Monte Carlo based uncertainty analysis method, to investigate the propagation of errors in the estimation of adsorption parameters on the performance variability of a glucose-fructose \textsc{smb} system.
Since discrepancies between simulation and experimental results have often been observed in the \textsc{smb} field, it is necessary to embrace a method that inherently considers uncertainties.

In this study, Bayesian inference will be used to investigate the uncertainty of operating conditions of a glucose-fructose \textsc{smb} system on the multiple performance indicators.
Formulation of Bayes theorem into the process design of \textsc{smb} processes will firstly be presented.
The glucose-fructose case study is chosen such that cross-validation with the algebraic solution can be implemented.
Bayesian inference allows for measurements of uncertainties in a system to be propagated in a mathematically consistent manner.
Bayesian inference differs with the commonly adopted frequentist methods in the chromatographic field (\ie, Fisher information based method and bootstrapping approach).
The Monte Carlo based methods belong to the bootstrapping approach.
Although large uncertainties of adsorption and experimental parameters have been reported, by using both bootstrap and Fisher information matrix, it is hardly presented that how the uncertainties of operating conditions affect the predictive behaviour of \textsc{smb} models from the Bayesian perspective.

The operating conditions of the four-zone \textsc{smb} unit will be optimized by a stochastic algorithm, Markov Chain Monte Carlo (\textsc{mcmc}), with respect to conflicting objectives (purity, yield as a demonstration and can be further generalized to other objectives).
Pareto fronts will be calculated for illustrating the best compromises between each pair of the performance indicators.
Unlike multi-objective optimization algorithms (\eg, non-dominated sorted genetic algorithm \citep{zhang2002multiobjective}, strength Pareto evolutionary algorithms) that try to eliminate all the non-dominated points during optimization, \textsc{mcmc} serves on the sampling purpose, which is interested in sampling the Pareto optimal points as well as those near Pareto optimal.
For the sampling purpose, \textsc{mcmc} not only accepts proposals with better objective value, but also accepts moves heading to non-dominated points with certain probability.
Since numerical solutions are not derived from the equilibrium theory (thus not limited to linear scenarios), they are capable to enumerate, in principle, all the infinite combinations of operating conditions to depict the complete separation region for nonlinear situations.
By using the \textsc{mcmc} sampling, however, a coherent connection between numerical solutions and the triangle theory can be generated; the complete separation region of the triangle theory can be intrinsically sketched for the \textsc{smb} processes described by any kind of adsorption isotherms.

\section{Theory}

\subsection{Column model}
The general rate model (\textsc{grm}) that accounts for various levels of mass transfer resistance in phases \citep{guiochon2006fundamentals} is used to describe the transport behaviour of components in the columns.
Convection and axial dispersion in the bulk liquid are considered, as well as film mass transfer and pore diffusion in the porous beads:

\begin{subequations}\label{eq:GRM}
    \begin{align}
    \frac{\partial c_i^j}{\partial t} & = -u_\text{int}^j \frac{\partial c_i^j}{\partial z} + D_{\text{ax}}^j \frac{\partial^2 c_i^j}{\partial z^2} - \frac{1-\varepsilon_c}{\varepsilon_c} \frac{3}{r_p} k_{f,i}^j \left(c_i^j - c_{p,i}^j(r\!=\!r_p)\right) \\
    \frac{\partial c_{p,i}^j}{\partial t} & = D_{p,i}^j \left(\frac{\partial^2 c_{p,i}^j}{\partial r^2} + \frac{2}{r}\frac{\partial c_{p,i}^j}{\partial r}\right) - \frac{1-\varepsilon_p}{\varepsilon_p}\frac{\partial q_i^j}{\partial t}
    \end{align}
\end{subequations}
In Eq.~\eqref{eq:GRM}, $z \in [0,L]$ denotes the axial position where $L$ is the column length, while $r \in [0,r_p]$ denotes the radial position where $r_p$ is the particle radius.
Furthermore, $c_i^j$, $c_{p,i}^j$ and $q_i^j$ denote the interstitial, stagnant and stationary phase concentrations of component $i\in \{1,\dots,M\}$ in column $j \in \{1, \dots, N\}$, respectively.
$t$ is the time, $\varepsilon_c$ and $\varepsilon_p$ are the column and particle porosities, $u_{\text{int}}^j$ interstitial velocity; $D_{\text{ax}}^j$ is the axial dispersion coefficient, $D_{p,i}^j$ the effective pore diffusion coefficient, and $k_{f,i}^j$ the film mass transfer coefficient.
At the column inlet and outlet, Danckwerts boundary conditions \citep{Barber1998Boundary} are applied:
\begin{equation}
    \begin{cases}
        \left. \dfrac{\partial c_i^j}{\partial z} \right|_{z=0} &= \dfrac{u_{\text{int}}^j}{D_{\text{ax}}^j} \left(c_i^j(z\!=\!0) - c_{\text{in},i}^j \right) \\
        \left. \dfrac{\partial c_i^j}{\partial z} \right|_{z=L} &= 0
    \end{cases}
    \label{eq:Danckwerts_column}
\end{equation}
where $c_{\text{in},i}^j$ is the inlet concentration of component $i$ in column $j$ (cf.~Eq.~\eqref{eq:node_balance}).
The boundary conditions at the particle surface and centre are described by:

\begin{equation}
    \begin{cases}
        \left. \dfrac{\partial c_{p,i}^j}{\partial r} \right|_{r=r_p} &= \dfrac{k_{f,i}^j}{\varepsilon_p D_{p,i}^j} \left(c_i^j - c_{p,i}^j (r\!=\!r_p)\right) \\
        \left. \dfrac{\partial c_{p,i}^j}{\partial r} \right|_{r= 0} &= 0
    \end{cases}
    \label{eq:Danckwerts_surface}
\end{equation}

When the adsorption equilibrium can be described by the linear isotherm, the concentration of each component in the mobile phase of column $j$, $c_{p,i}^j$, and in the particle phase of column $j$, $q_i^j$, are linearly correlated:
\begin{equation}
    q_i^j = H_i c_{p,i}^j
    \label{eq:linear}
\end{equation}

\subsection{Node model}
In this study, two adjacent columns ($j$, $j\!+\!1$) are connected via a node $j$.
Therefore, the node $j$ is located at the downstream side of column $j$ and the upstream side of column $j\!+\!1$.
Only one or none of feed (F), desorbent (D), raffinate (R), or extract (E) streams exists at a time for a node.
The occasion that nodes are not connected to a port (\ie, in the interior of a zone) occurs when more columns than zones are present, such as, eight columns in a four-zone scheme.
A circular \textsc{smb} loop is closed when the column indices follows $\mathrm{mod}(j,N) = 1$ (\ie, by identifying column $j=N\!+\!1$ with column $j=1$).
The inlet concentration of component $i$ in column $j\!+\!1$ is calculated from mass balance of the node $j$:

\begin{align}
    c_{\text{in},i}^{j+1} = \frac{c_{\text{out},i}^j Q^j + \delta_i^j}{Q^{j+1}}
    \label{eq:node_balance}
\end{align}
where $c_{\text{out},i}^j = c_i^j(t, z\!=\!L)$ denotes the outlet concentration of component $i$ in column $j$, $Q^j = \varepsilon_c u_{\text{int}}^j \pi d_c^2/4$ the zonal flowrates and $d_c$ the column diameter.
The $\delta^j$ is determined by the current role of node $j$ (\ie, F, D, R, E or none):

\begin{align}
    \delta^j_i = \left\{  \begin{array}{l@{\quad \quad}l}
        \phantom{-}c_{\text{in,i}}^F Q^F & \text{feed} \\
        \phantom{-}c_{\text{in,i}}^D Q^D & \text{desorbent} \\
        -c_{\text{out,i}}^j Q^R & \text{raffinate} \\
        -c_{\text{out,i}}^j Q^E & \text{extract} \\
        \phantom{-}0       & \text{none}
    \end{array} \right.
    \label{eq:delta}
\end{align}
where $c_{\text{in},i}^F$ and $c_{\text{in},i}^D$ are the component concentrations at feed and desorbent ports, and $Q^F$, $Q^D$, $Q^R$, $Q^E$ the volumetric flowrates at the feed, desorbent, raffinate, and extract ports.
Column shifting is implemented by periodically permuting $\delta$ each switching time $t_s$.
From a mathematical point of view, the switching operation can be represented by a shifting of the initial and boundary conditions for the single column \citep{klatt2002model}.

\subsection{Performance indicators}
In this study, performance indicators are all defined in terms of components, $i \in \{1, \dots, M\}$, withdrawn at a four-zone \textsc{smb} node, $j \in \{E, R\}$, within one switching time $t_s$.
The definitions of performance indicators are all based on concentration integrals $\bar{c}_{\text{out},i}^j$ of component $i$ at node $j$ upon cyclic steady state (\textsc{css}), averaged over one switching time, $t_s$:
\begin{equation}
    \bar{c}_{\text{out},i}^j = \frac{1}{t_s} \int_{t=(k-1)t_s}^{k t_s} c_i^j(t, z\!=\!L)\, \dd t
    \label{eq:switching}
\end{equation}
In Eq.~\eqref{eq:switching}, $k > k_\text{CSS}$ denotes the switching number.

Purity, yield and productivity are commonly used performance indicators for process assessment.
The purity of a component $i$, $\mathtt{Pu}_{i}^j$, is the average concentration of this component in the collected fraction of all components at port $j$, Eq.~\eqref{eq:purity}. 
The yield, $\mathtt{Y}_i^j$, is the ratio between the amount of the desired component $i$ in the collected fraction at port $j$ and the amount injected in the column through the feed port, Eq.~\eqref{eq:yield}.
The productivity, $\mathtt{Pr}_{i}^{j}$, is the withdrawn mass of the component $i$ at port $j$ per collection time relative to the total volume of the utilized packed bed in all columns, Eq.~\eqref{eq:prod}.
\begin{equation}
    \mathtt{Pu}_{i}^j = \frac{\bar{c}_{\text{out},i}^j }{\sum\limits_{k=1}^{M} \bar{c}_{\text{out},k}^j}
    \label{eq:purity}
\end{equation}
\begin{equation}
    \mathtt{Y}_i^j = \frac{Q^j \bar{c}_{\text{out},i}^j}{Q^F c_{\text{in},i}^F}
    \label{eq:yield}
\end{equation}
\begin{equation}
    \mathtt{Pr}_{i}^{j} = \frac{Q^j \bar{c}_{\text{out},i}^j}{(1-\varepsilon_c) V_c\, N}
    \label{eq:prod}
\end{equation}
In addition, throughput, solvent consumption, pressure drops and cycle times can also be considered to assess performance.

\subsection{Multi-objective optimization}{\label{moo}}

Consider a \textsc{smb} model with $M$ components and $n$ parameters, $\theta \in \mathbb{R}^n$.
Generally, operating conditions of \textsc{smb} processes are systematically optimized by numerical algorithms such that the above performance indicators ($\Psi = [\mathtt{Pu}_i^j, \mathtt{Y}_i^j, \mathtt{Pr}_i^j, \dots]$) are all maximized.
However, there are trade-off relationships among the performance indicators.
For example, maximizing $\mathtt{Pu}_i^j$ would result in low values of $\mathtt{Y}_i^j$ and $\mathtt{Pr}_i^j$, and vice versa.
Therefore, multi-objective optimization is applied.

A set of objectives can be combined into a single objective by adding each objective a pre-multiplied weight (the weighted method \citep{marler2010weighted}), or keeping just one of the objectives and with the rest of the objectives constrained (the $\varepsilon$-constraint method \citep{chankong2008multiobjective,mavrotas2009effective}).
The latter method is used in this work; as a demonstration, maximizing the yields of components at the corresponding withdrawn ports with their purities constrained to be larger than thresholds $\varepsilon_i^j$:
\begin{equation}
    \begin{array}[ ]{r l}
        \min        & f(\theta) = - \displaystyle \sum_{j\in\{R,E\}} \mathtt{Y}_i^{j} \\
        \text{s.t.} & \begin{cases}
            c^j(\theta) : \mathtt{Pu}^j_i - \varepsilon_i^j \geqslant 0 \\
            \theta_{\min} \leqslant \theta \leqslant \theta_{\max}
        \end{cases}
    \end{array}
    \label{eq:eps_constraint}
\end{equation}
The searching domain is bounded, $[\theta_{\min}, \theta_{\max}]$.
Extension of Eq.~\eqref{eq:eps_constraint} to ternary or quaternary objectives is straightforward.
The inequalities $c^j(\theta),\, j \in \{R,E\}$ is lumped into the objective function using penalty terms in this study, such that it can be solved as a series of unconstrained minimization problems with increasing penalty factors, $d_k$:
\begin{equation}
    \min \mathcal{H}(\theta; d_k) = f(\theta) + d_k g(\theta)
    \label{eq:objective}
\end{equation}
In Eq.~\eqref{eq:objective}, the penalty function is chosen as $g(\theta) = \sum_{j\in\{R,E\}} \norm{ \min\{0, c^j(\theta)\} }^2$.

\subsection{Bayesian inference}
In the Bayesian framework, inference conclusions are made in terms of probabilities, which is used as the fundamental measure of uncertainties.
Four distributions are briefly introduced here and will be discussed in detail later:
\begin{enumerate}
    \item Prior distribution, $p(\theta)$, represents a population of possible parameter values. It is where we can express our knowledge about parameters on the inference. The prior distribution should include all possible values of $\theta$, but realistically the information about $\theta$ contained in the data will far outweigh reasonable probability distributions.
    \item Likelihood function (also called sampling distribution), $p(\Psi|\theta)$, is a conditional distribution that describes the probability of a data set, $\Psi$, for given parameters $\theta$.
        It is typically formulated as a function of the model parameters, and contains all the information relative to the evaluation of statistical evidence.
        The sampling distribution plays a major role in Bayesian inference over the prior distribution.
    \item Marginal distribution, $p(\Psi)$, is an integral of the joint distribution (\ie, the product of prior distribution and likelihood function) over the parameter space. 
        It describes the probability of data set, $\Psi$, that is a constant independent of model and model parameters.
        However, the constant value is very hard to compute.
    \item Posterior distribution, $p(\theta|\Psi)$, is a conditional distribution that describes the probability of the parameter set, $\theta$, given the data $\Psi$.
        We also refer to the posterior distribution as the target distribution.
        It contains the desired information on the sought parameters.
\end{enumerate}
The above distributions are related to each other by the Bayes theorem \citep{gelman2014bayesian}:

\begin{equation}
    p(\theta|\Psi)=\frac{p(\Psi|\theta) p(\theta)}{p(\Psi)} 
    \label{eq:bayes}
\end{equation}
where $p(\Psi) = \int p(\Psi, \theta)\, \dd \theta  = \int p(\theta) p(\Psi|\theta)\, \dd \theta$.
The integral is very hard and computationally expensive to calculate for multi-dimensional distributions.
Therefore, Bayesian inference is often realized by approximating the unnormalized posterior distribution (cf.~Eq.~\ref{eq:likelihood}) via sampling (\ie, \textsc{mcmc}).
Technically, the \textsc{mcmc} sampling allows to calculate the constant value, $p(\Psi)$.
However, this is pointless, as 1) it is sufficient to sample from Eq.~\ref{eq:likelihood} and 2) integrating $p(\Psi)$ takes the similar computational effort with calculating the sought posterior distribution, $p(\theta|\Psi)$.

\begin{equation}
    p(\theta|\Psi) \propto p(\Psi|\theta) p(\theta) 
    \label{eq:likelihood}
\end{equation}

Different types of methods can be applied to solve the minimization of $\mathcal{H}(\theta;d_k)$, such as deterministic methods and heuristic methods. 
Bayesian inference shall be adopted in this study.
In order to use Bayesian inference, the $\mathcal{H}(\theta;d_k)$ is further formulated as a likelihood function in the following exponential form:
\begin{equation}
    p(\Psi|\theta) \overset{\text{def}}= \exp\pbk{ - \frac{1}{2} \mathcal{H}(\theta; d_k) }
    \label{eq:likelihood2}
\end{equation}

\subsection{Markov Chain Monte Carlo}
Based on the Markov chain theory to generate chains, \textsc{mcmc} is able to sample from complicated distributions.
The more states that are collected, the more closely the distribution of the samples matches the desired distribution.
Various \textsc{mcmc} algorithms have been developed, which mainly differ in computational complexity, robustness, and speed of convergence.

\subsubsection{Metropolis algorithm}\label{sec:MetropolisAlgo}
Metropolis algorithm \citep{metropolis1953equation} is one of the blocking bricks.
It is a random-walk algorithm with Gaussian proposal for sampling the operating parameters $\theta$.
The algorithm proceeds as follows:

\begin{enumerate}
    \item Initialize a starting point, $\theta^0$, for example, from the prior distribution; construct a covariance matrix, $\Sigma$, for the proposal distribution (Gaussian distribution in the present work).
    \item For $k = 1, 2, \dots$:
        \begin{itemize}
            \item Based on the previous sample, $\theta^k$, a candidate $\tilde{\theta}$ is drawn from the Gaussian proposal distribution, $\mathcal{N}(\theta^k, \Sigma)$.
            \item A ratio $\gamma(\tilde{\theta}, \theta^k)$ of posterior distributions of the candidate, $\tilde{\theta}$, and the previous sample, $\theta^k$, with respect to the desired target distribution is calculated:
\begin{equation}\begin{split}
    \gamma(\tilde{\theta}, \theta^k) &= \frac{p(\tilde{\theta}|\Psi)}{p(\theta^k|\Psi)} = \frac{p(\Psi|\tilde{\theta}) p(\tilde{\theta})}{p(\Psi|\theta^k) p(\theta^k)} \\ 
    & = \exp\cbk{ -\frac{1}{2} \pbk{ \mathcal{H}(\tilde{\theta}; d_k) - \mathcal{H} (\theta^k; d_k)} } \frac{p(\tilde{\theta})}{p(\theta^k)}
    \label{eq:ratio}
\end{split}\end{equation}

            \item The candidate is conditionally accepted with the following probability, where the random number, $\beta$, is drawn from the uniform distribution on the interval $[0,1]$.
        \begin{equation}
            \theta^{k+1} = \left\{  \begin{array}{l@{\quad}l}
            \tilde{\theta} & \beta \leqslant \min\pbk{1, \gamma(\tilde{\theta}, \theta^k)} \\
            \theta^k & \text{otherwise} \end{array} \right.
        \end{equation} 

            \item The index $k$ is increased by one and the procedure is repeated until a stopping criterion is satisfied (\eg, a predefined number of samples is reached).
        \end{itemize}
\end{enumerate}

Eventually, a sequence of random samples whose distribution approximates the target density is obtained.
In implementation, a portion of samples (\eg, \SI{25}{\percent}) are discarded as \textit{burn-in} to diminish the influence of the starting point, $\theta^0$.
Though the Metropolis algorithm has simple and easy to implement features, it has low efficiency.
In sampling, numerous candidates can be rejected, resulting slow convergence to the target distribution.
Further, when one chain is trapped into a local mode, it might never converge to the target density.
Hence, several enhancements have been proposed in the literature.
In this study, an adaptive Metropolis strategy and a delayed rejection \citep{haario2006dram} are applied to alleviate the drawbacks.

\subsubsection{Adaptive Metropolis strategy}

The convergence of the Metropolis algorithm can be accelerated by adapting shape of the proposal distribution (\ie, Gaussian distribution in this study), as determined by the covariance matrix, $\Sigma$.
Fisher information matrix can be a typical choice for the initial covariance matrix, $\Sigma_0$,
\begin{equation}
    \Sigma_0 = \tilde{\sigma}_0\,  V (S^T S)^{-1} V^T 
    \label{eq:sigma_initial}
\end{equation}
where the matrices $U$, $S$ and $V$ are the singular value decomposition of the Jacobian of the chromatography model, \ie, sensitivity of the chromatogram with respect to the parameters at $\theta^0$.
$\tilde{\sigma}_0$ is a hyperparameter.
Further details can be found in \ref{app:covariance}. 
Another practical option is to run a pre-simulation beforehand and then calculate an approximated $\Sigma_0$ from the samples.

Initially, iterations of the Metropolis algorithm are performed with $\Sigma_0$ for a fixed and pre-defined number.
Then, the covariance matrix $\Sigma$ is adapted in regular intervals, based on the history of the Markov chain

\begin{equation}
\Sigma =  c\, \text{Cov}(\Theta) + \varepsilon_a I \label{eq:sigma_metro_adaptive}
\end{equation}
where $\Theta \in \mathbb{R}^{k\times n}$ is the single Markov chain at iteration $k$.
$I$ is the identity matrix, and $c$ is a scaling factor proposed by \citet{gelman2014bayesian} to be $c = 2.4^2 / \sqrt{n}$.
A small $\varepsilon_a > 0$ prevents $\Sigma$ from becoming singular.
The covariance matrix $\text{Cov}(\Theta)$ is calculated by using all previously computed samples of the current Markov chain.

\subsubsection{Delayed rejection}

The efficiency can be enhanced by delaying the rejection of candidates.
Instead of discarding a proposal when $\gamma(\tilde{\theta}, \theta^k) \leqslant 1$ is satisfied, a next stage of the Metropolis algorithm is performed with a shrunken covariance matrix, $a \Sigma$.
A shrinking factor of $a=0.1$ can be applied.

Delayed rejection is applied to increase the robustness and efficiency of the adaptive Metropolis strategy.
The initial covariance matrix $\Sigma_0$ often can not be approximated correctly, which leads to a high rejection rate at the beginning.
Consequently, this makes the adaption very slow since only few distinct points are available for estimating a better covariance matrix.
In order to evolve the proposal covariance matrix to the structure of target density, more points need to be accepted, by taking another more cautious move from the starting point when the first move is rejected (possibly due to the inappropriate large scaling).
Further details can be found in \citep{haario2006dram} and \ref{app:dr}.

\subsubsection{Convergence criteria and effective sample size}

Consider that the samples collected from $m$ multiple chains are denoted as $\Phi \in \mathbb{R}^{k\times n\times m}$; samples for an estimated parameter $\theta_\ell\, (\ell \in \{1,\dots,n\})$ are labelled as $\Phi^\ell_{\varkappa, \iota}, (\varkappa\in\{1,\dots,k\}, \iota \in\{1, \dots, m\}$).
Stopping criteria are required to stop the \textsc{mcmc} simulations.
A potential scale reduction factor (a.k.a.~Gelman criteria \citep{gelman2014bayesian}) can be applied to $\Phi^\ell$ to assess convergence conditions.
It is a square root of the ratio of sample variances, 
\begin{equation}
    \widehat{R}_\ell = \sqrt{\frac{\widehat{\text{var}}^{+}(\mathcal{W}, \mathcal{B})}{\mathcal{W}}}
    \label{eq:gelman}
\end{equation}
which is based on between- and within-chain variances:
\begin{equation}
    \mathcal{B} = \frac{k}{m-1} \sum_{\iota=1}^m \left( \bar{\Phi}_{.\iota}^\ell - \bar{\Phi}_{..}^\ell \right)^2
    \label{eq:B}
\end{equation}
\begin{equation}
    \mathcal{W} = \frac{1}{m} \sum_{r=1}^n s_\iota^2
    \label{eq:W}
\end{equation}
Note that the \emph{burn-in} part has been discarded in length of the chains $k$ here.
$\bar{\Phi}_{.\iota}^\ell = \frac{1}{k} \sum_{\varkappa=1}^k \Phi_{\varkappa, \iota}^\ell$ is the mean value of the chain $\iota$.
$\bar{\Phi}_{..}^\ell$ is the mean value of the mean vector of $m$ chains, $\frac{1}{m} \sum_{\iota=1}^m \bar{\Phi}_{.\iota}^\ell$.
Thus, $\mathcal{B}$ defines the between-chain variance; $s_\iota^2$ denotes the within-chain variance of the chain $\iota$.
The sample variance, $\widehat{\text{var}}^{+}(\mathcal{W}, \mathcal{B})$, is estimated by a weighted average of $\mathcal{W}$ and $\mathcal{B}$, namely
\begin{equation}
    \widehat{\text{var}}^{+}(\mathcal{W}, \mathcal{B}) = \frac{k-1}{k} \mathcal{W} + \frac{1}{k} \mathcal{B}
\end{equation}

Upon convergence of the \textsc{mcmc} algorithm diagnosed by, for instance, the above criteria, the samples collected, since then, from the multiple chains can be mixed up to approximate the target distribution.
The effective number of independent simulation draws (\ie, \emph{effective sample size}) for any optimized parameter $\theta_\ell$ can be estimated from Eq.~\ref{eq:eff}.
\begin{equation}
    n_{\text{eff}} = \frac{mk}{1 + 2 \displaystyle\sum_{t=1}^{\infty} \rho_t}
    \label{eq:eff}
\end{equation}
where $\rho_t$ is the autocorrelation of the mixed-up chain of the parameter $\theta_\ell$ at lag $t$; 
In practice, however, we barely have a finite simulation length, so the calculation has to be approximated.

\section{Case}\label{sec:case}

A monosaccharide mixture of glucose and fructose on a laboratory scale four-zone \textsc{smb} process is used as a model example in this study, $i\in \{\text{glc, fru}\}$ \citep{klatt2002model}.
There are eight columns and two columns in each zone.
The plant was reported to be operated at \SI{60}{\celsius}.
The liquid density can be considered as constant for the given feed concentration, and the adsorption isotherm is well-described by Henry's law.
The parameters are classified into two catalogues, $[\varphi, \theta]$; $\theta$ is treated as decision variables (\ie, degree of freedom) and shall be optimized in this study while $\varphi$, as experimental conditions, is kept unchanged and directly used, see Tab.~\ref{tab:literature_data}.

\begin{table}
    \centering
    \scriptsize
    \caption{Reference parameters for the separation of glucose and fructose on an 8-column \textsc{smb} laboratory plant.}
    \label{tab:literature_data}
    \begin{tabular}{c c l S c}
        \toprule
        Catalogue   &       Symbol          &   Description             &  {Value}          &   Unit \\   
        \midrule
        \multirow{5}{*}{$\varphi$} &           $d_c$           &   column diameter         & 2.6e-2            &   \si{\metre} \\
                    &   $d_p$           &   particle diameter       & 3.25e-3           &   \si{\metre} \\
                    &   $\varepsilon$   &   column void             & 0.38              &   \\  
                    &   $H_i$           &   Henry constants         & {$[0.28, 0.54]$}  &   \\
                    &   $c_i^F$         &   Feed concentration      & 3.05e3            &   \si{\mole\per\cubic\metre}\\
        \midrule
        \multirow{7}{*}{$\theta$} &            $L$             &   column length           & 5.36e-1           &   \si{\metre} \\
                    &   $t_s$           &   Switching time          & 1.552e3           &   \si{\second} \\
                    &   $Q^\text{rec}$  &   Recycle flowrate        & 1.395e-7          &   \si{\cubic\metre\per\second}\\  
                    &   $Q^F$           &   Feed flowrate           & 2.00e-8          &   \si{\cubic\metre\per\second} \\
                    &   $Q^R$           &   Raffinate flowrate      & 2.66e-8          &   \si{\cubic\metre\per\second} \\ 
                    &   $Q^D$           &   Desorbent flowrate      & 4.14e-8          &   \si{\cubic\metre\per\second} \\ 
                    &   $Q^E$           &   Extract flowrate        & 3.48e-8          &   \si{\cubic\metre\per\second} \\ 
        \bottomrule
    \end{tabular}
\end{table}

In this study, all columns are assumed to be identical (\eg, packing density and porosities) and initially empty.
Pure buffer is used as the desorbent, \ie, $c_{\text{in},i}^D = \SI{0}{\mole\per\cubic\metre}$.
The feed concentrations are converted from $c_{\text{in},i}^F = \SI{550}{\gram\per\litre}$, assuming that fructose and glucose have the same molar mass of $\SI{180.16}{\gram\per\mol}$.
Four typical axial dispersion coefficients $D_\text{ax}^j,\, j\in\{\text{I}, \text{II}, \text{III}, \text{IV}\}$, are used since the values are not given in \citet{klatt2002model}.
The volumetric flowrates $Q^j$ in zones $j\in\{\text{II}, \text{III}, \text{IV}\}$ are calculated from the recycle and F, D, E, R flowrates.
The interstitial velocities in zones are calculated from the respective volumetric flowrates, $u^j_{\text{int}} = Q^j / (\varepsilon_c A)$, where $A$ is the cross-section area.

The equilibrium-dispersive model (\textsc{edm}) was used in \citet{klatt2002model}, as the axial dispersion effects can not be neglected and the mass transfer are fast but not infinitely fast.
Though the \textsc{cadet} is originally designed for solving the \textsc{grm}, it can be adapted for solving the \textsc{edm} as follows: 
Eq.~\ref{eq:GRM} is spatially discretized using the finite volume method with only one radial cell, $N_r = 1$ (the axial column dimension is discretized into $N_z = 40$ cells).
A very small particle porosity is used, $\varepsilon_p = \num{1e-5}$, such that the column porosity is asymptotic to the total porosity, $\lim_{\varepsilon_p \rightarrow 0} \varepsilon_c = \lim_{\varepsilon_p \rightarrow 0} \frac{\varepsilon_t - \varepsilon_p}{1 - \varepsilon_p} = \varepsilon_t$. 
The effective pore diffusion coefficient and the film mass transfer coefficient are made sure not to be rate limiting, $D_p = \SI{5e-5}{\square\metre\per\second}$ and $k_f = \SI{1.6e4}{\metre\per\second}$.

The mathematical models described above for each column of the \textsc{smb} processes are weakly coupled together and then iteratively solved.
The open-source code has been published on Github, \url{https://github.com/modsim/CADET-SMB.git}.
\textsc{cadet-smb} repeatedly invokes \textsc{cadet} kernel to solve each individual column model.
The resulting system of ordinary differential equations is solved using an absolute tolerance of $\num{1e-10}$, relative tolerance of $\num{1e-6}$, an initial step size of $\num{1e-14}$ and a maximal step size of $\num{5e6}$.
\textsc{cadet} is also an open-source software published on Github, \url{https://github.com/modsim/CADET.git}.
All numerical simulations are computed on an Intel(R) Xeon(R) system with 16 CPU cores (64 threads) running at \SI{2.10}{\giga\hertz}.


A stochastic multi-objective sampling algorithm, \textsc{mcmc}, is applied in this study to optimize the operating conditions.
The Metropolis algorithm, incorporating with delayed rejection and adjusted Metropolis, has been published as open-source software on Github, \url{https://github.com/modsim/CADET-MCMC.git}.
Non-informative prior distribution $p(\theta)$ is used.
Samples are collected until either the Gelman criteria for each parameter, $\widehat{R}_\ell,\, \ell\in\{1,\dots,n\}$, is smaller than $\num{1.1}$ or the maximal sample size (\ie, \num{400}) is reached.
$\widehat{R}_\ell$ declines to 1 when $k\rightarrow \infty$; but we generally have been satisfied with setting 1.1 as a threshold.
Smooth probability density of each marginal posterior distribution is estimated by a MATLAB routine, \texttt{ksdensity}.
Multiple chains of \textsc{mcmc} are used, but not intrinsically.
Two \textsc{mcmc} simulation instances (\ie, $m = 2$) are run simultaneously on the computing node; meanwhile, samples of each chain are written into the shared memory of the node.
An additional program of the convergence diagnose is running parallel on the node, accessing the convergence conditions periodically with the samples having been stored on the shared memory.
The Pareto fronts in this study describe two-dimensional trade-offs between a pair of the performance indicators, $\Psi$.
Thus, the non-dominated stable sort method of Pareto front is applied to generate the frontiers \citep{duh2012learning}.

\section{Results and discussion}

\subsection{Multi-objective optimization}

\begin{table}
    \centering
    \scriptsize
    \caption{Boundary conditions of the operating parameters of the four-zone scheme.}
    \label{tab:domain}
    \begin{tabular}{c l S S S}
        \toprule
        {\multirow{2}{*}{Symbol}} & {\multirow{2}{*}{Description}} & \multicolumn{2}{c}{{Value}} & {\multirow{2}{*}{Unit}} \\
        \cline{3-4}
                                  &                                &   {min}       & {max}       & \\ 
        \midrule                                                   
        $L$                       & column length                  & 50.0e-2       & 60.0e-2     & \si{\metre} \\
        $t_s$                     & switching time                 & 1.5e3         & 1.6e3       & \si{\second} \\
        $Q^{\text{rec}}$          & recycle flowrate               & 1.0e-7        & 1.8e-7      & \si{\cubic\metre\per\second} \\
        $Q^{F}$                   & feed flowrate                  & 1.5e-8        & 2.5e-8      & \si{\cubic\metre\per\second} \\
        $Q^{D}$                   & desorbent flowrate             & 3.5e-8        & 4.5e-8      & \si{\cubic\metre\per\second} \\
        $Q^{E}$                   & extract flowrate               & 3.0e-8        & 4.0e-8      & \si{\cubic\metre\per\second} \\
        \bottomrule
    \end{tabular}
\end{table}
Searching domain of the operating parameters, $\theta$, is listed in Tab.~\ref{tab:domain}.
The boundary intervals are based on the optimal condition of \citet{klatt2002model} with additional safety margins.
The maximal sampling length of \textsc{mcmc} is $k = 400$ with the \emph{burn-in} length of 50.
The autocorrelation plot for each parameter, $\theta_\ell,\, \ell\in\{1,\dots,n\}$, is shown in the \ref{app:autocorr}.
The Gelman criteria for the optimized parameters are $\widehat{R} = [1.07, 0.99, 1.01, 0.99, 1.00, 0.99]$ when $k = 309$.
As seen from the convergence diagnose (\ie, $\widehat{R}_\ell = 1.07$), the parameter, column length, converges slower than the other parameters.
The average effective sample size of all parameters upon convergence $n_{\text{eff}} = \num{78}$ in this case.
Therefore, no more than $n_{\text{eff}}$ samples are required to approximate the target distribution.
The sampling length and the effective sample size are rather small in the framework of Bayesian inference.
This is partially because of that the initial point used for the \textsc{mcmc} sampling is a \emph{optimal} point located in the stationary region of the multivariate posterior distribution; partially it is a four-zone \textsc{smb} with the linear isotherm.
If an initial point that is far away from the stationary region was used, it could exert significant impact on the efficiency and convergence of the \textsc{mcmc} algorithm.

\begin{figure}
    \centering
    \begin{subfigure}{0.49\textwidth}{\includegraphics[width=\textwidth]{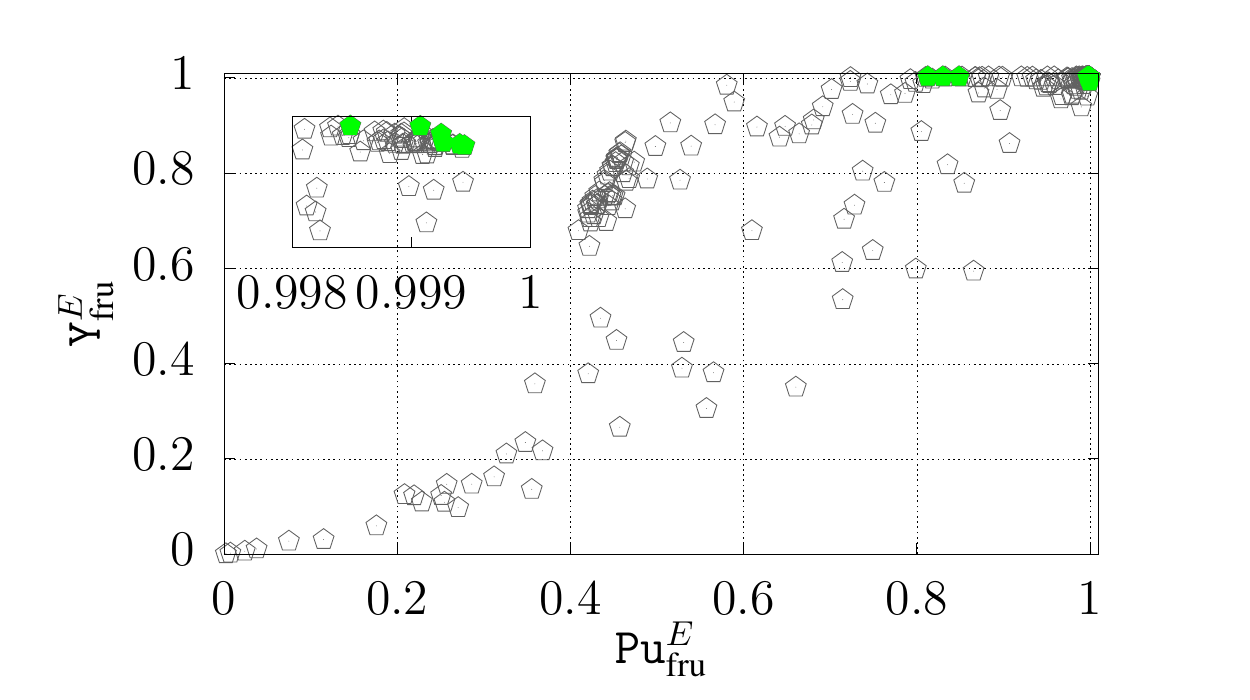}} \caption{} \label{fig:pareto_a} \end{subfigure}
    \begin{subfigure}{0.49\textwidth}{\includegraphics[width=\textwidth]{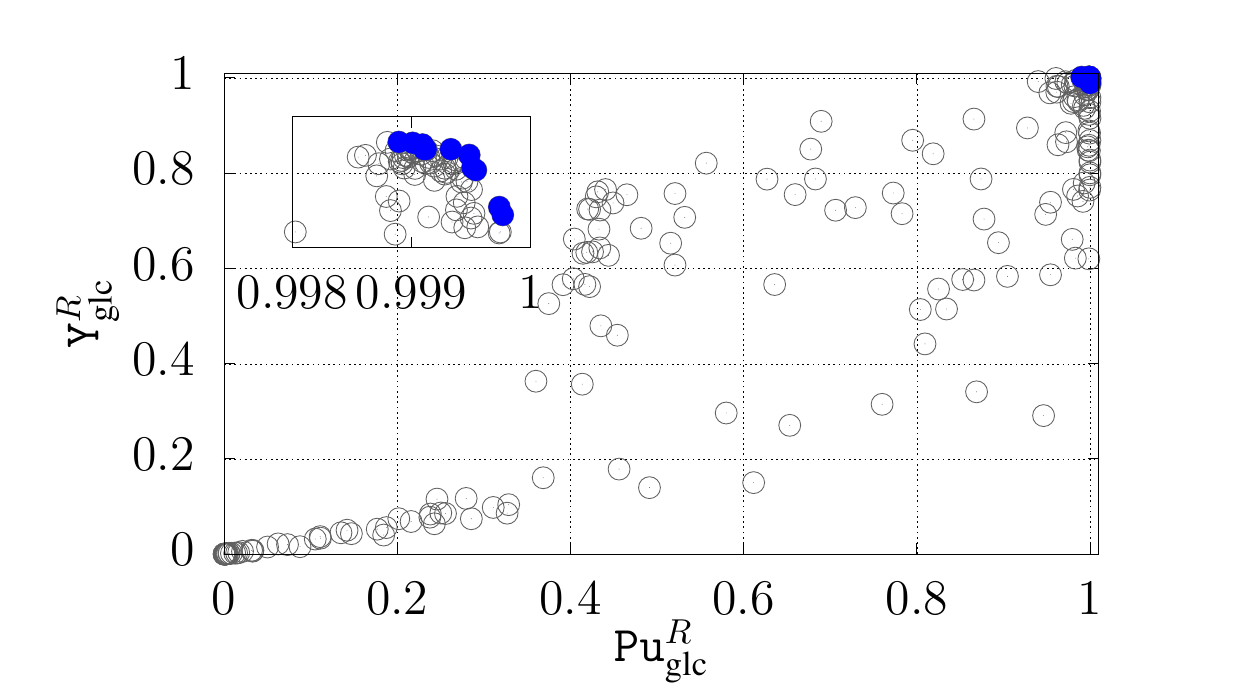}} \caption{} \label{fig:pareto_b} \end{subfigure}
    \begin{subfigure}{0.49\textwidth}{\includegraphics[width=\textwidth]{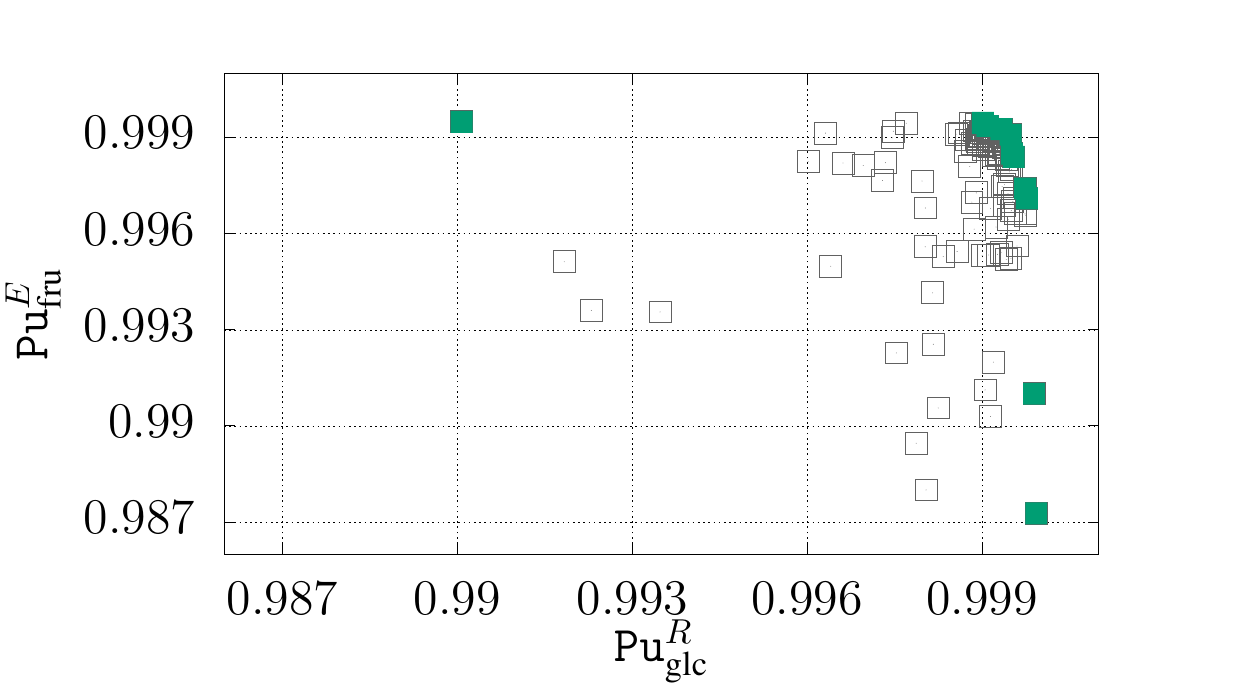}} \caption{} \label{fig:pareto_c} \end{subfigure}
    \begin{subfigure}{0.49\textwidth}{\includegraphics[width=\textwidth]{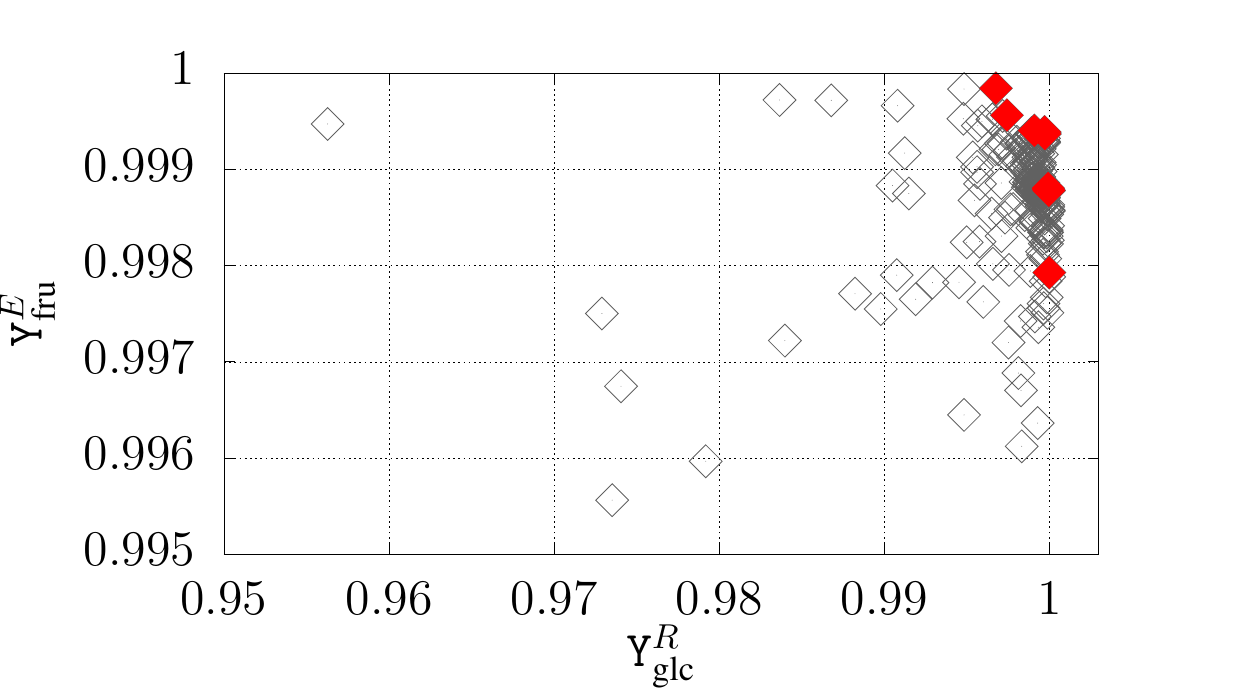}} \caption{} \label{fig:pareto_d} \end{subfigure}
    \begin{subfigure}{0.49\textwidth}{\includegraphics[width=\textwidth]{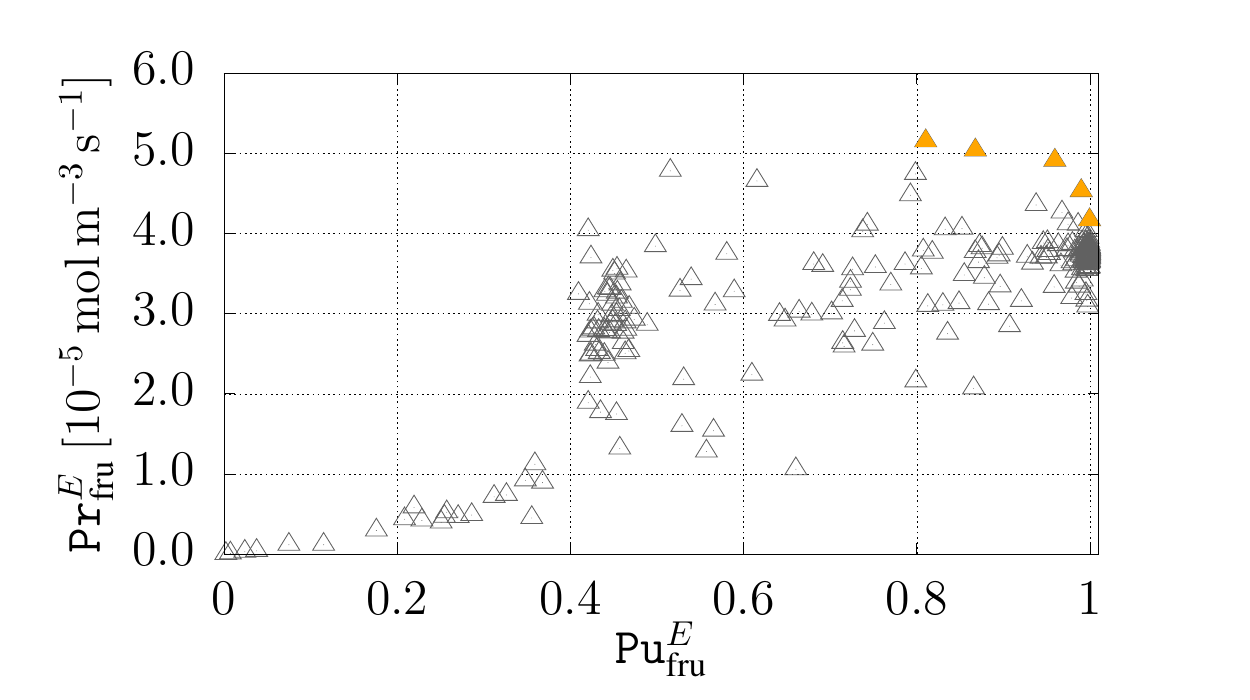}} \caption{} \label{fig:pareto_e} \end{subfigure}
    \begin{subfigure}{0.49\textwidth}{\includegraphics[width=\textwidth]{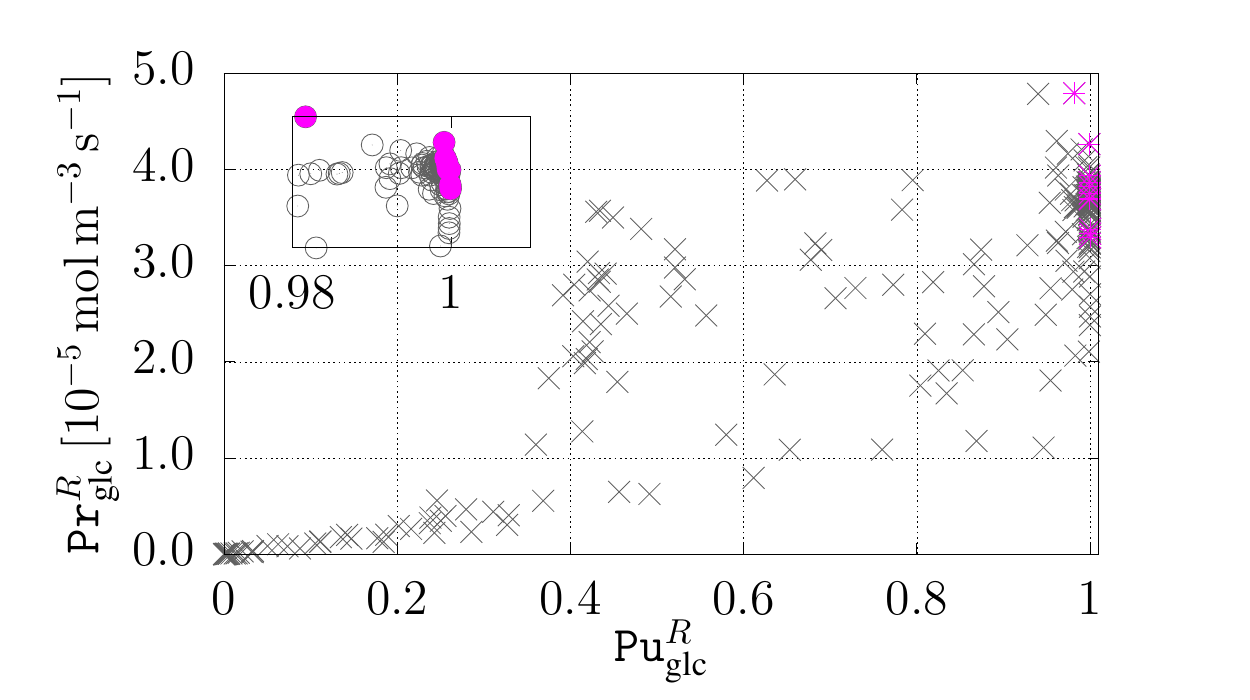}} \caption{} \label{fig:pareto_f} \end{subfigure}
    \caption{Pareto fronts between each pair of performance indicators $\Psi = [\mathtt{Pu}_i^j, \mathtt{Y}_i^j, \mathtt{Pr}_i^j],\, i\in\{\text{glc, fru}\},\, j\in\{R,E\}$. The coloured symbols illustrate the Pareto fronts, while the gray symbols illustrate samples from \textsc{mcmc} and show the convergence trajectories. $a, b, c$ are three points chosen for exemplification.}
    \label{fig:paretos}
\end{figure}

As the glucose has lower value of Henry coefficient than fructose $H_{\text{glc}} < H_{\text{fru}}$ (thus lower retention time), it is collected at the raffinate port of the four-zone \textsc{smb} process; while fructose is collected at the extract port.
Fig.~\ref{fig:paretos} shows the Pareto fronts between each pair of the performance indicators $\Psi = [\mathtt{Pu}_i^j, \mathtt{Y}_i^j, \mathtt{Pr}_i^j]$ considered in the case study.
At the extract port, high yield of fructose $\mathtt{Y}_{\text{fru}}^E$ can be achieved with a wide range of purity $\mathtt{Pu}_{\text{fru}}^E$ from $60\%$ to $100\%$ (Fig.~\ref{fig:pareto_a});
While at the raffinate port, rather high purity of glucose, $\mathtt{Pu}_{\text{glc}}^R = 99.9\%$, can be achieved within a wide range of yield, $\mathtt{Y}_{\text{glc}}^R$ (Fig.~\ref{fig:pareto_b}).
It implies that the less adsorbed component, glucose, is inclined to spread to the extract port, resulting in low purity of fructose at the extract port, and low yield of glucose at the raffinate port. 
This shall be explained from another point of view in \emph{section}~\ref{sec:triangle}. 
With resort to the four-zone \textsc{smb} scheme, operating conditions that render high purities $[\mathtt{Pu}_{\text{glc}}^R, \mathtt{Pu}_{\text{fru}}^E]$ and yield $[\mathtt{Y}_{\text{glc}}^R, \mathtt{Y}_{\text{fru}}^E]$ at both outlet streams can be found (cf.~Fig.~\ref{fig:pareto_c}-\ref{fig:pareto_d}).
According to the definitions of yield and productivity, both indicators increase with amounts of the product collected.
However, productivity can also be enhanced by reducing the switching time, $t_s$, that is optimized.
At a rather high purity requirement of $\mathtt{Pu}_{\text{fru}}^E = 99.9\%$, a productivity $\mathtt{Pr}_{\text{fru}}^E$ of ca.~\SI{4.0e-2}{\mole\per\cubic\metre\per\second} can be achieved at the extract port; while at purity of $80\%$, the productivity increases to ca.~\SI{5.0e-2}{\mole\per\cubic\metre\per\second} (Fig.~\ref{fig:pareto_e}).
At the raffinate port, high productivity of $\mathtt{Pr}_{\text{glc}}^R = \SI{5.0e-2}{\mole\per\cubic\metre\per\second}$ at hight purity of $\mathtt{Pu}_{\text{glc}}^R = 99.9\%$ can be achieved (Fig~\ref{fig:pareto_f}).

\begin{figure}
    \centering
    \includegraphics[width=0.6\textwidth]{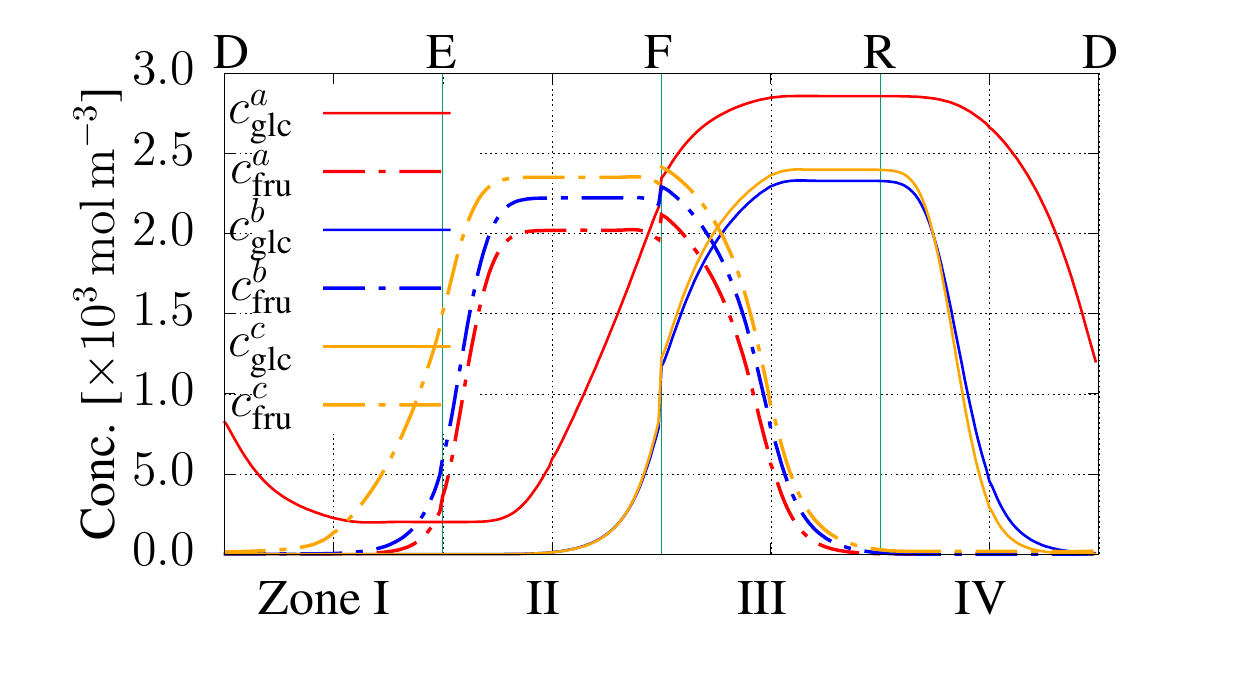}
    \caption{Chromatograms of the characteristic points ($a,b,c$) on the Pareto fronts.  D, E, F, R denote desorbent, extract, feed and raffinate respectively.}
    \label{fig:chromas}
\end{figure}

The Pareto fronts provide full trade-off information of the four-zone \textsc{smb} scheme for separating glucose and fructose.
Three characteristic points on the Pareto frontiers, as listed in Tab.~\ref{tab:cpoint}, are exemplified.
Points, $a, b, c$, were chosen on purpose; $\mathtt{Pu}_{\text{fru}}^E$ is increasing while $\mathtt{Pu}_{\text{glc}}^R$ is decreasing.
On the language of triangle theory, operating points are picked from the pure raffinate region ($a$), the complete separation region ($b$) and pure extract region ($c$), respectively.
The corresponding chromatograms along the four-zone \textsc{smb} train are shown in Fig.~\ref{fig:chromas}.
As stated in \emph{section} \ref{moo}, the multi-objective optimization can be straightforwardly generalize to other performance indicators, beyond the demonstration of yield and purity here.

\begin{table}
    \centering
    \scriptsize
    \caption{Performance indicators of the three characteristic points on the Pareto fronts}
    \begin{tabular}{c c c c c S S}
        \toprule
        Point  & $\mathtt{Pu}_{\text{fru}}^E$ $[\%]$  & $\mathtt{Pu}_{\text{glc}}^R$ $[\%]$ & $\mathtt{Y}_{\text{fru}}^E$  & $\mathtt{Y}_{\text{glc}}^R$  &  {$\mathtt{Pr}_{\text{fru}}^E$ [\si{\mole\per\cubic\metre\per\second}]}   &  {$\mathtt{Pr}_{\text{glc}}^R$ [\si{\mole\per\cubic\metre\per\second}]} \\
        \midrule
        $a$     &   $85.21$   &   $99.98$   &   1.00  &   0.83      &   4.06e-2  &   3.35e-2 \\
        $b$     &   $99.92$   &   $99.92$   &   1.00  &   1.00      &   3.74e-2  &   3.74e-2 \\
        $c$     &   $99.95$   &   $99.01$   &   0.99  &   1.00      &   4.16e-2  &   4.21e-2 \\
        \bottomrule
    \end{tabular}
    \label{tab:cpoint}
\end{table}

\subsection{Uncertainty of operating conditions}

\begin{figure}
    \centering
    \begin{subfigure}{0.4\textwidth}{\includegraphics[width=\textwidth]{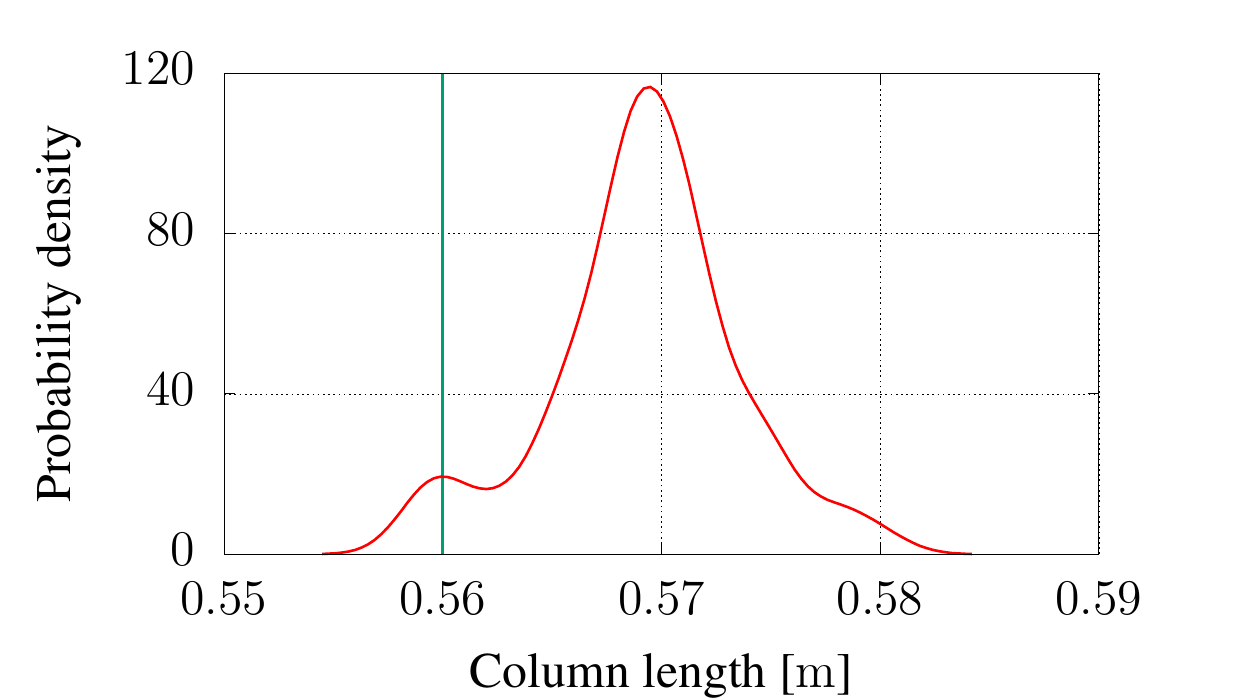}} \caption{Column length} \label{fig:postDis_a} \end{subfigure}
    \begin{subfigure}{0.4\textwidth}{\includegraphics[width=\textwidth]{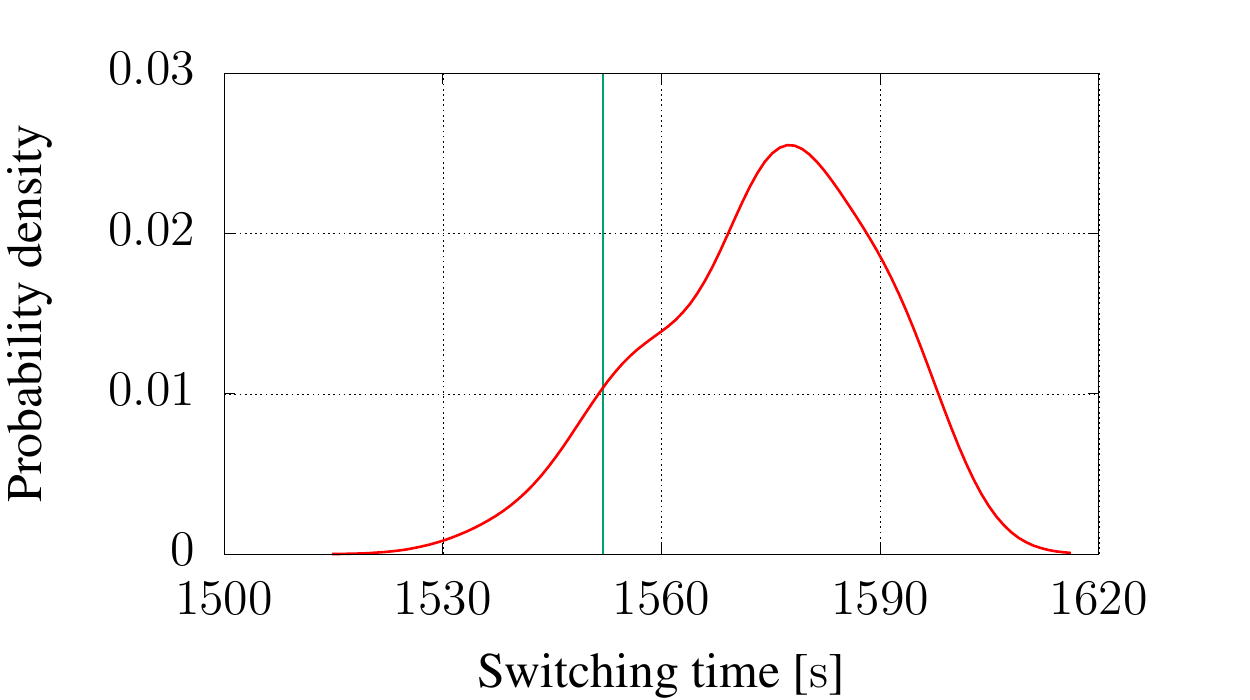}} \caption{Switching time} \label{fig:postDis_b} \end{subfigure}
    \begin{subfigure}{0.4\textwidth}{\includegraphics[width=\textwidth]{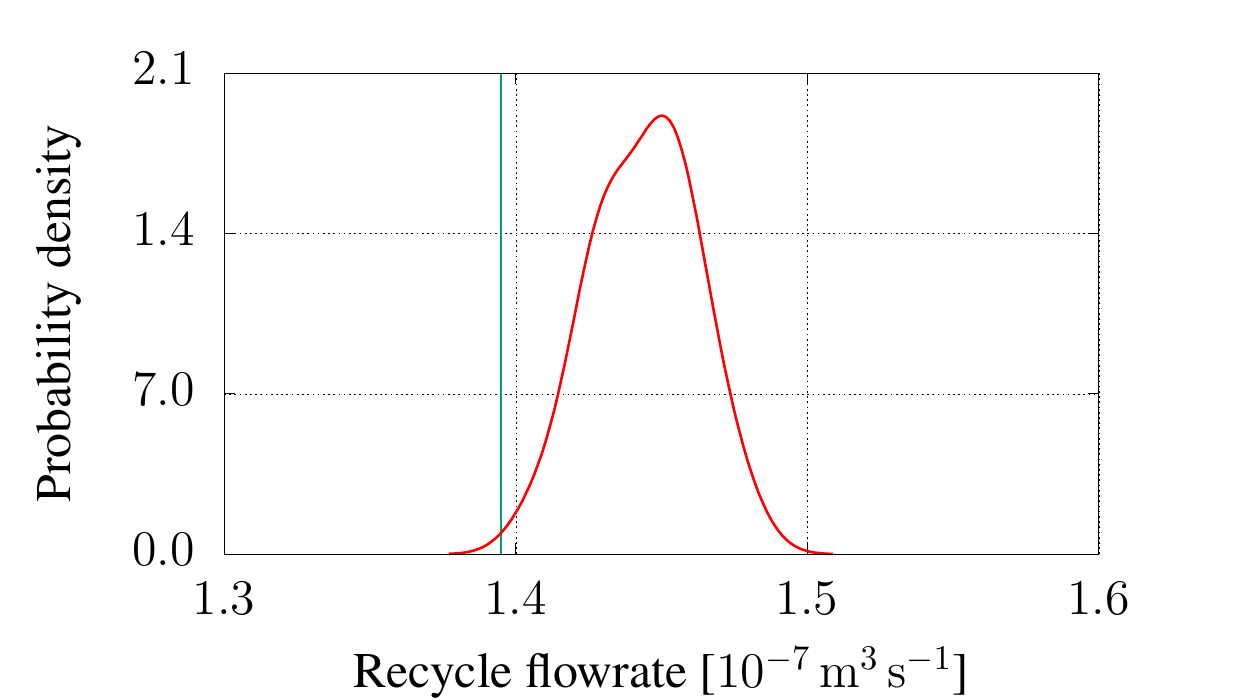}} \caption{Recycle flowrate} \label{fig:postDis_c} \end{subfigure}
    \begin{subfigure}{0.4\textwidth}{\includegraphics[width=\textwidth]{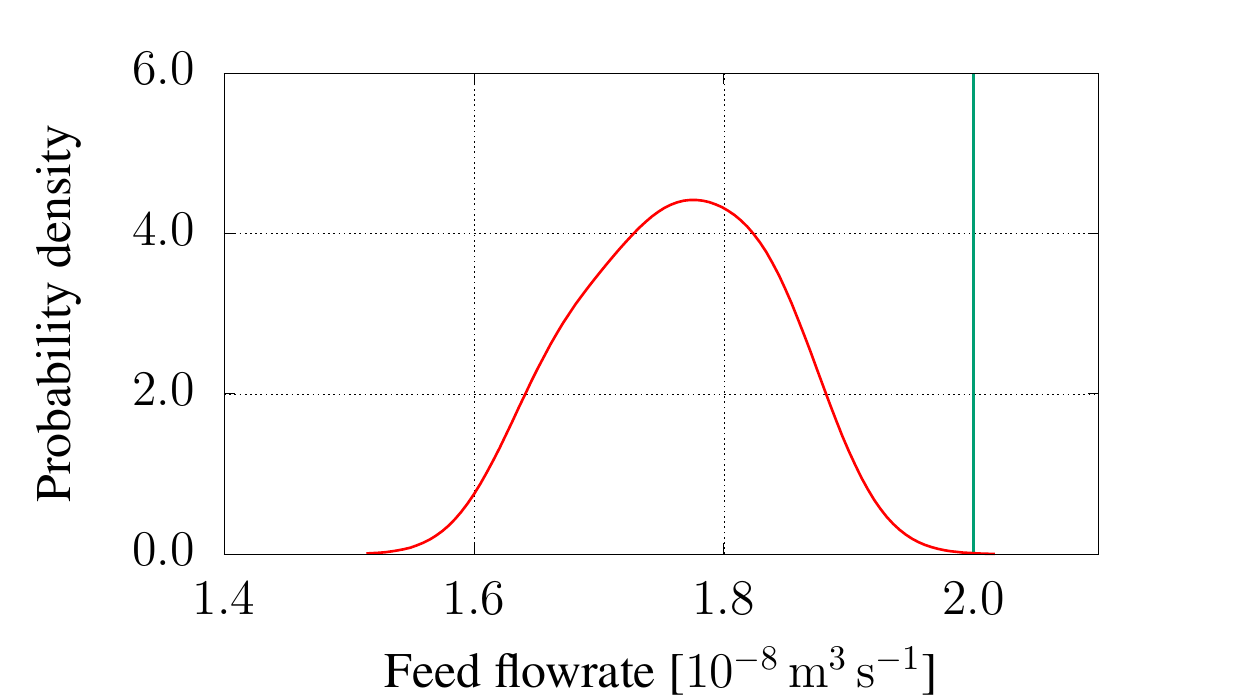}} \caption{Feed flowrate} \label{fig:postDis_d} \end{subfigure}
    \begin{subfigure}{0.4\textwidth}{\includegraphics[width=\textwidth]{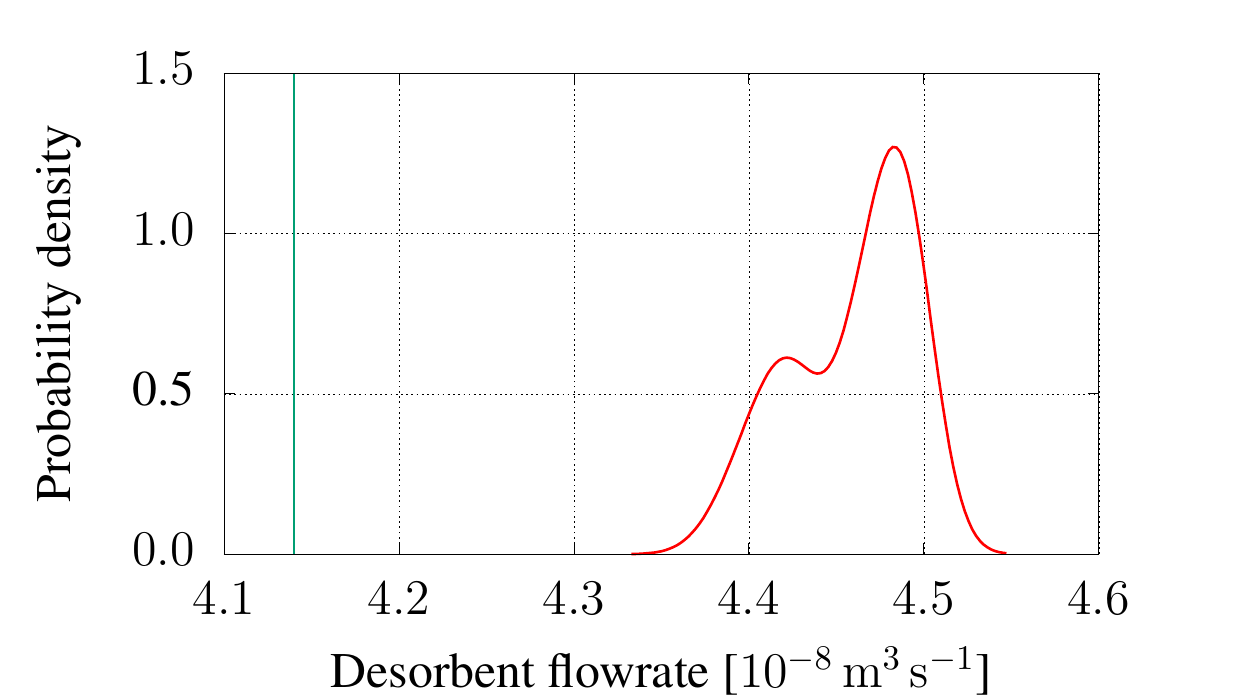}} \caption{Desorbent flowrate} \label{fig:postDis_e} \end{subfigure}
    \begin{subfigure}{0.4\textwidth}{\includegraphics[width=\textwidth]{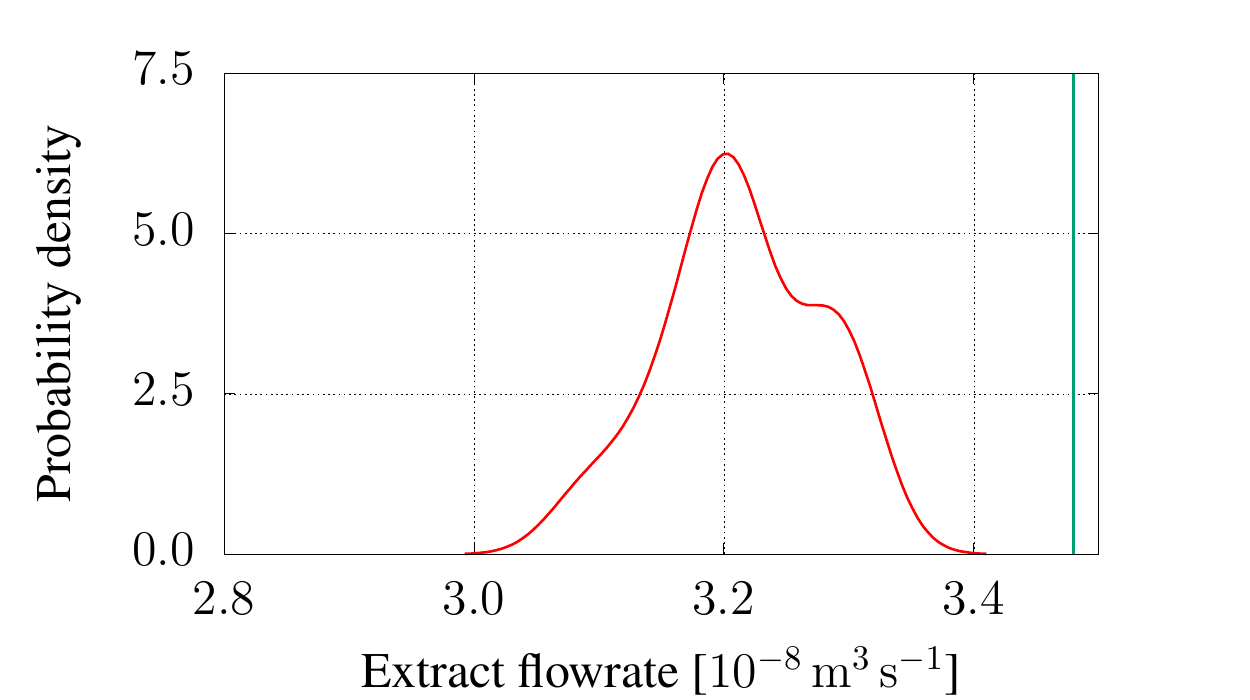}} \caption{Extract flowrate} \label{fig:postDis_f} \end{subfigure}
    \begin{subfigure}{0.4\textwidth}{\includegraphics[width=\textwidth]{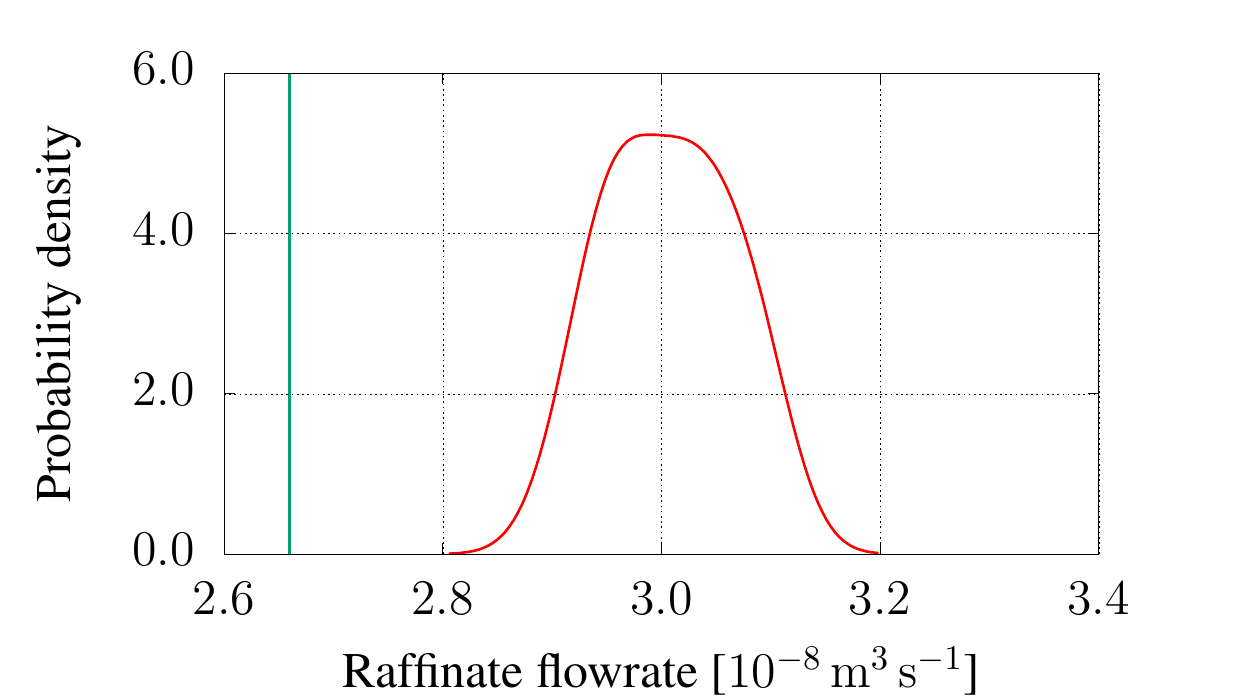}} \caption{Raffinate flowrate} \label{fig:postDis_g} \end{subfigure}
    \begin{subfigure}{0.4\textwidth}{\includegraphics[width=\textwidth]{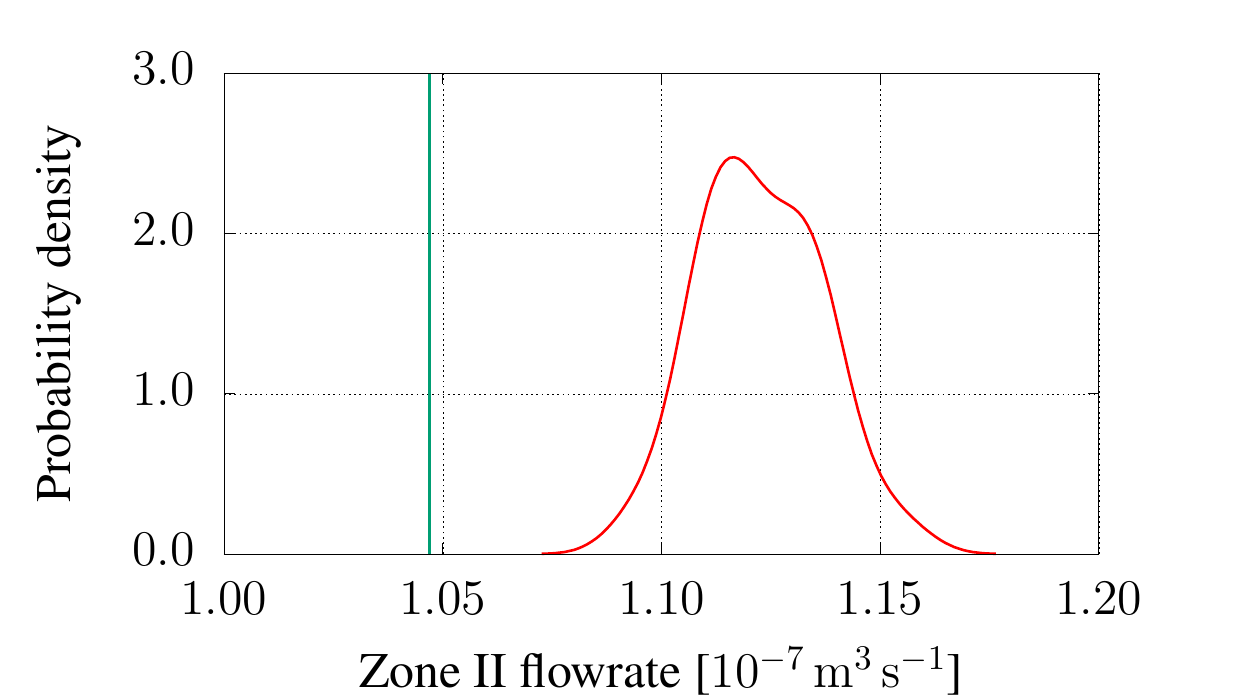}} \caption{Zone II flowrate} \label{fig:postDis_h} \end{subfigure}
    \begin{subfigure}{0.4\textwidth}{\includegraphics[width=\textwidth]{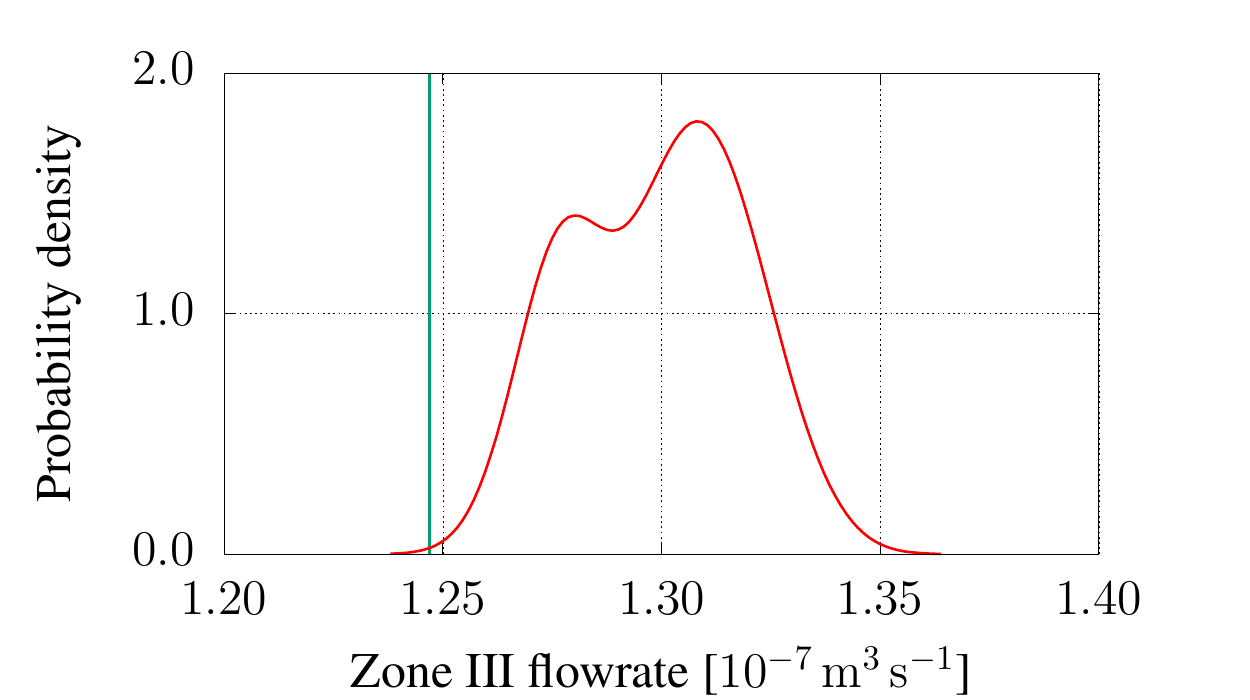}} \caption{Zone III flowrate} \label{fig:postDis_i} \end{subfigure}
    \begin{subfigure}{0.4\textwidth}{\includegraphics[width=\textwidth]{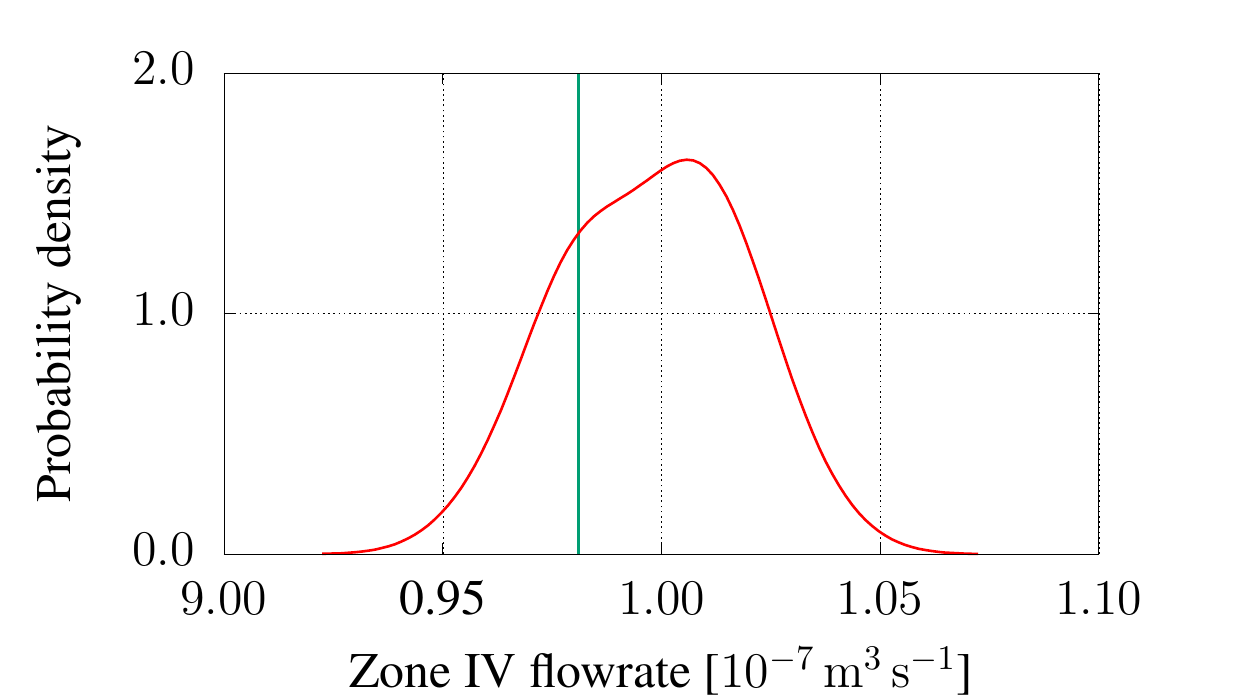}} \caption{Zone IV flowrate} \label{fig:postDis_j} \end{subfigure}
    \caption{Posterior distributions of the operating conditions of the four-zone \textsc{smb} process, which result in the $\mathtt{Y}_i^j = 1$, $\mathtt{Pu}_i^j \geqslant 99.9\%$ target. The blue lines denote the parameter values from \citet{klatt2002model}; purple lines for the point $a$, black lines for the point $b$, yellow lines for the point $c$.}
    \label{fig:postDis}
\end{figure}

Marginal posterior distributions of the operating conditions listed in Tab.~\ref{tab:domain} were generated directly from the collected \textsc{mcmc} samples; while the distributions of $Q^R, Q^{\text{II}}, Q^{\text{III}}, Q^{\text{IV}}$ were calculated with the following vector operations:
\begin{equation}
    \begin{split}
    Q^{\text{II}} & = Q^{\text{rec}} - Q^E \\
    Q^{\text{III}} & = Q^{\text{II}} + Q^F \\
    Q^{\text{IV}} & = Q^{\text{rec}} - Q^D \\
    Q^{R} & = Q^D + Q^F- Q^E \\
    \label{eq:vector}
    \end{split}
\end{equation}
Fig.~\ref{fig:postDis_a}-\ref{fig:postDis_j} show the marginal posterior distributions, resulting in the $\mathtt{Y}_i^j = 1$, $\mathtt{Pu}_i^j \geqslant 99.9\%$ target.
They are all unimodal.
The widths of the unimodal distributions indicate that the operating conditions are well-determined on the parameter estimation point of view.

For the flowrates of recycle, $Q^{\text{rec}}$, feed, $Q^F$ and raffinate, $Q^R$, they have distributions of smooth Gaussian shapes; while for the rest parameters, they are asymmetric (\eg, Fig.~\ref{fig:postDis_e}, \ref{fig:postDis_i}), skewed (\eg, Fig.~\ref{fig:postDis_h}, \ref{fig:postDis_j}) and tailing (\eg, Fig.~\ref{fig:postDis_b}, \ref{fig:postDis_f}).
In other words, they might be rough in shapes.
The roughness could potentially be polished by running the \textsc{mcmc} simulations much longer such that each convergence diagnostics goes to 1 (\ie, $k\rightarrow \infty, \widehat{R}_\ell \rightarrow 1$).
But, the roughness does not substantially disturb to make inference here, as the convergence criteria are satisfied with rather small thresholds and samples larger than the effective sample size have been collected.
Roughness of distribution shapes and computational consumption are under a trade-off relationship.
In \textsc{smb}, \textsc{css} can be obtained only in an asymptotic sense, which means that it will take a large number of periods, hence a long time to enter into it.
It takes $\num{80}\sim\num{144}$ iterations (\ie, $\num{1.7}\sim\SI{3.5}{\hour}$ on the computing node) for each \textsc{smb} simulation to converge to \textsc{css} with tolerance criterion of \num{1e-5}.
Longer sampling length of \textsc{mcmc} results in large computational burden.

The blue lines in Fig.~\ref{fig:postDis} are the parameter values from \citet{klatt2002model}.
As shown, the column length increases from \SI{0.56}{\meter} to \SI{0.57}{\metre}, correspondingly, the switching time increases from \SI{1552}{\second} to \SI{1580}{\second}.
The volumetric flowrates at feed and extract ports reduce, while all the other flowrates increase.
For comparison, the parameter sets of points ($a, b, c$) are also marked with coloured lines.

The distribution results are further analysed by credible intervals (CI), $[\breve{\delta}, \hat{\delta}]$, of each parameter $\theta_\ell$ (cf.~Eq.~\eqref{eq:CI}).
It is approximated by taking the $\alpha/2$ and $1 - \alpha/2$ percentiles from the collected samples while ignoring all other entries in the parameter vector.
\begin{equation}
    \begin{split}
        \int_{\text{CI}} p(\theta_\ell|y)\, \dd \theta_\ell & = 1 - \alpha, \quad0 < \alpha < 1\\
        p(\theta_\ell|y) & = \int \dots\! \int p(\theta|y)\, \dd (\theta_1,\dots,\theta_{\ell-1},\theta_{\ell+1},\dots, \theta_n)
        \label{eq:CI}
    \end{split}
\end{equation}
CI in the framework of Bayesian inference is equivalent to the confidence intervals in the frequentist statistics.
In Tab.~\ref{tab:CI}, the CI of operating conditions and the values with maximal posterior probabilities, $\mu$, are shown.
In this study, $1-\alpha = 0.66$, that is, $66\%$ credible intervals.
The credible intervals are, in majority, within $5\%$ deviation of $\mu$.
Therefore, the operating conditions are well-determined.

\begin{table}
    \centering
    \scriptsize
    \caption{Credible intervals of the operating conditions of the four-zone \textsc{smb} process.}
    \begin{tabular}{c S S S c c}
        \toprule
        $\theta$ & {$\mu$} & {$\breve{\delta}$} & {$\hat{\delta}$} & $\abs{\frac{\mu - \breve{\delta}}{\mu}}$ $\%$ & $\abs{\frac{\mu - \hat{\delta}}{\mu}}$ $\%$ \\
        \midrule
        $L$             &   0.569       & 0.566     & 0.573     & 0.5  & 0.7 \\
        $t_s$           &   1.577e3     & 1.558e3   & 1.592e3   & 1.2  & 1.0  \\ 
        $Q^{\text{rec}}$&   1.450e-7    & 1.430e-7  & 1.461e-7  & 1.4  & 2.1  \\
        $Q^F$           &   1.773e-8    & 1.669e-8  & 1.834e-8  & 5.9  & 3.4  \\
        $Q^D$           &   4.482e-8    & 4.420e-8  & 4.491e-8  & 1.4  & 0.2  \\
        $Q^E$           &   3.203e-8    & 3.157e-8  & 3.288e-8  & 1.4  & 2.7  \\   
        $Q^R$           &   2.988e-8    & 2.951e-8  & 3.069e-8  & 1.2  & 2.7  \\   
        $Q^{\text{II}}$ &   1.117e-7    & 1.110e-7  & 1.136e-7  & 0.6  & 1.7  \\       
        $Q^{\text{III}}$&   1.308e-7    & 1.277e-7  & 1.317e-7  & 2.4  & 0.7  \\           
        $Q^{\text{IV}}$ &   1.006e-7    & 9.805e-8  & 1.018e-7  & 2.5  & 1.2  \\       
        \bottomrule
    \end{tabular}
    \label{tab:CI}
\end{table}

\begin{figure}
    \centering
    \includegraphics[width=0.6\textwidth]{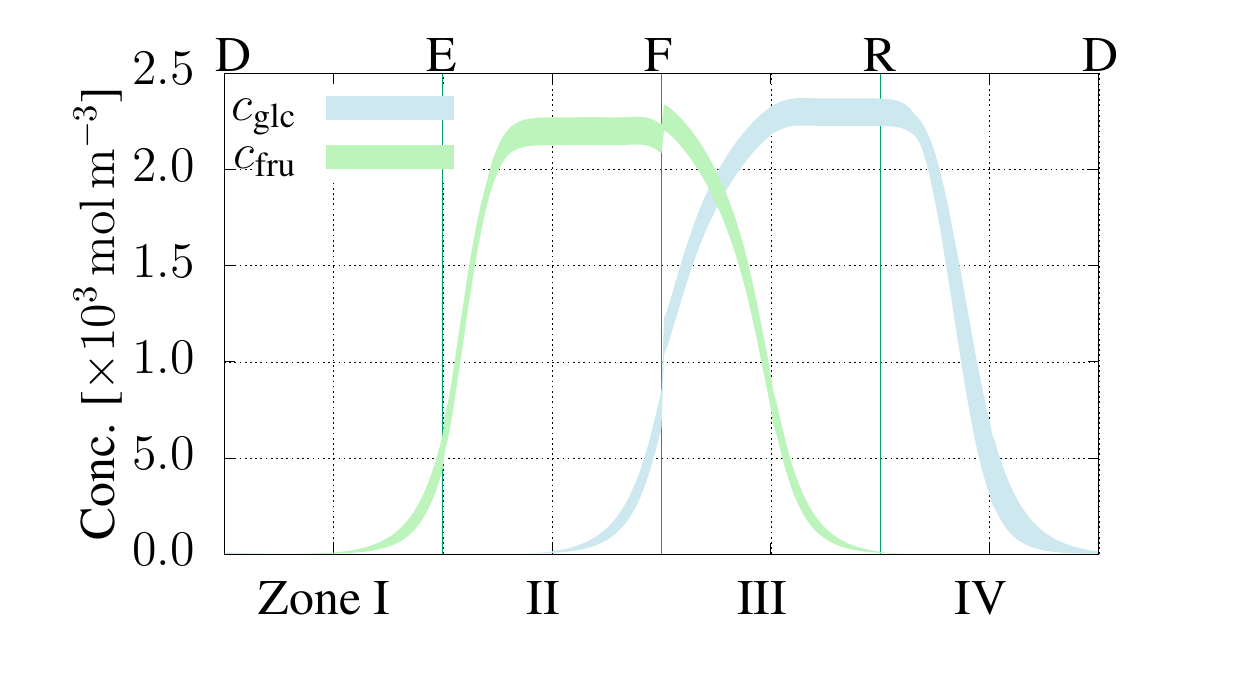}
    \caption{Posterior predictive check with the \textsc{mcmc} samples from the stationary region. D, E, F, R denote desorbent, extract, feed and raffinate respectively.}
    \label{fig:ppc}
\end{figure}

Posterior distributions of parameters can also be used to interpret the robustness of \textsc{smb} processes, by means of posterior predictive check (\textsc{ppc}) in the framework of Bayesian inference.
Only uncertainty of operating conditions are take into consideration in this study.
The effects caused by model selection, model and experimental calibrations have not been taken into account, such as feed concentration and composition, and poor estimation of kinetic parameters in the model calibration.
Fig.~\ref{fig:ppc} shows the \textsc{ppc} of resulting posterior distributions. 
\num{30} random sample from the posterior distributions were used to do forward simulation of the \textsc{smb} process.
Maximal and minimal chromatogram values at each observation point along the \textsc{smb} train were recorded.
In Fig.~\ref{fig:ppc}, the light blue filled curves indicate the deviation region of the chromatogram of glucose; while the light green ones show the deviation region of the chromatogram of fructose.
The \textsc{smb} process is robust, as it has tolerance to perturbations in the deviations of the posterior distributions (cf.~Fig.~\ref{fig:postDis}), though the highest concentration values vary.
\textsc{ppc} should be adopted as the \emph{chromatograms} for experimental data to compare, rather than a single chromatogram from a single set of operating condition.
Therefore, by considering uncertainties in process design and \textsc{ppc} of posterior distributions, discrepancies between experimental and simulated data can be explained.

\subsection{Mapping to the triangle theory}\label{sec:triangle}

\begin{figure}
    \centering
    \begin{subfigure}{0.5\textwidth}{\includegraphics[width=\textwidth]{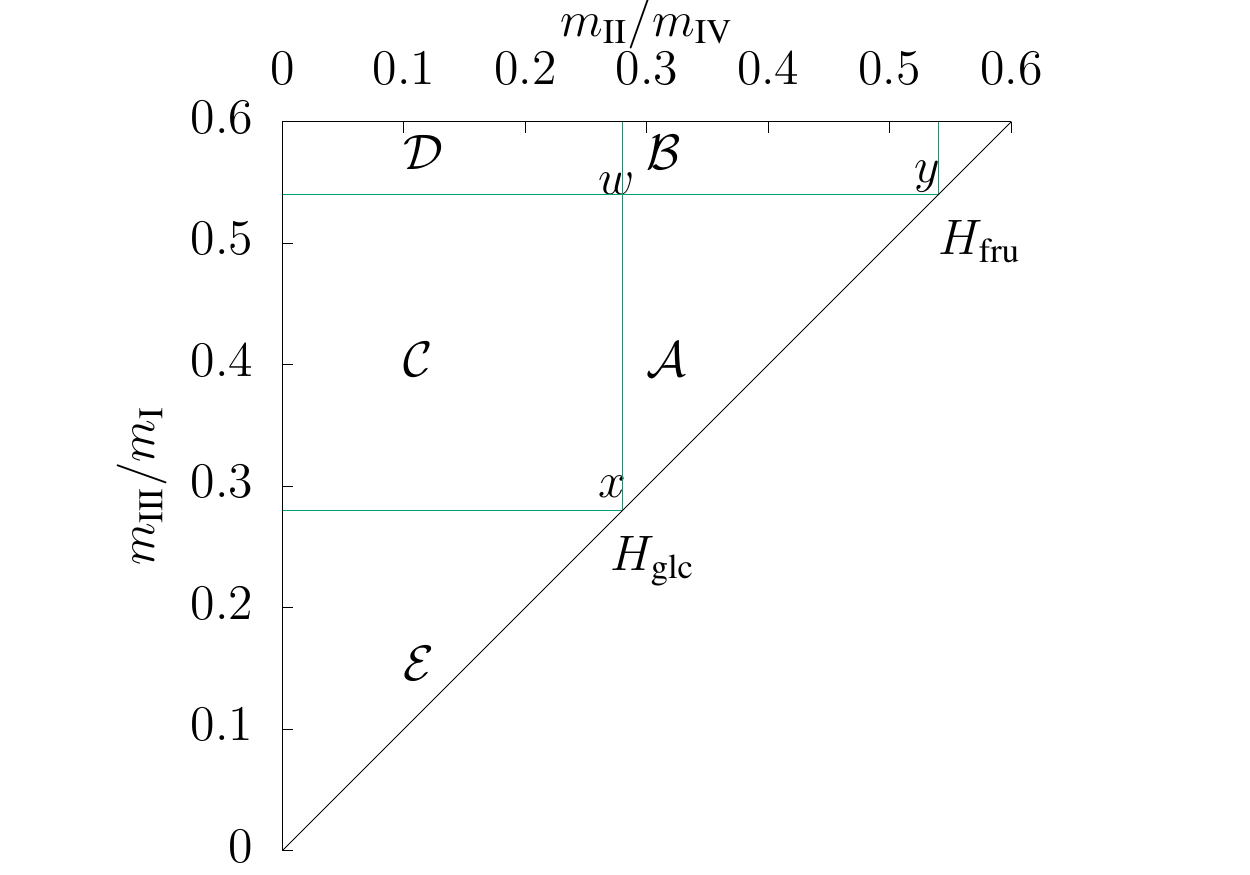}} \caption{} \label{fig:triangle_a} \end{subfigure}
    \begin{subfigure}{0.5\textwidth}{\includegraphics[width=\textwidth]{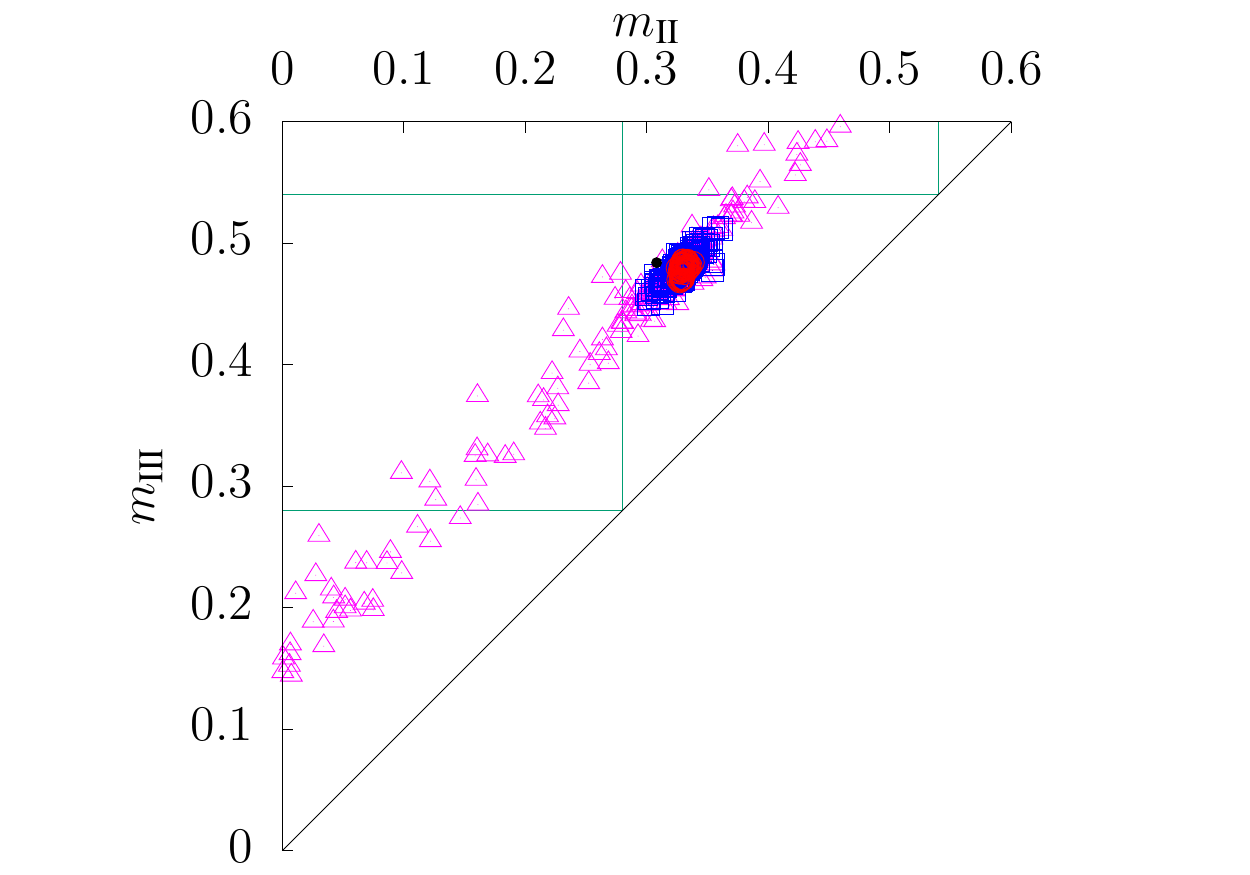}} \caption{} \label{fig:triangle_b} \end{subfigure}
    \begin{subfigure}{0.49\textwidth}{\includegraphics[width=\textwidth]{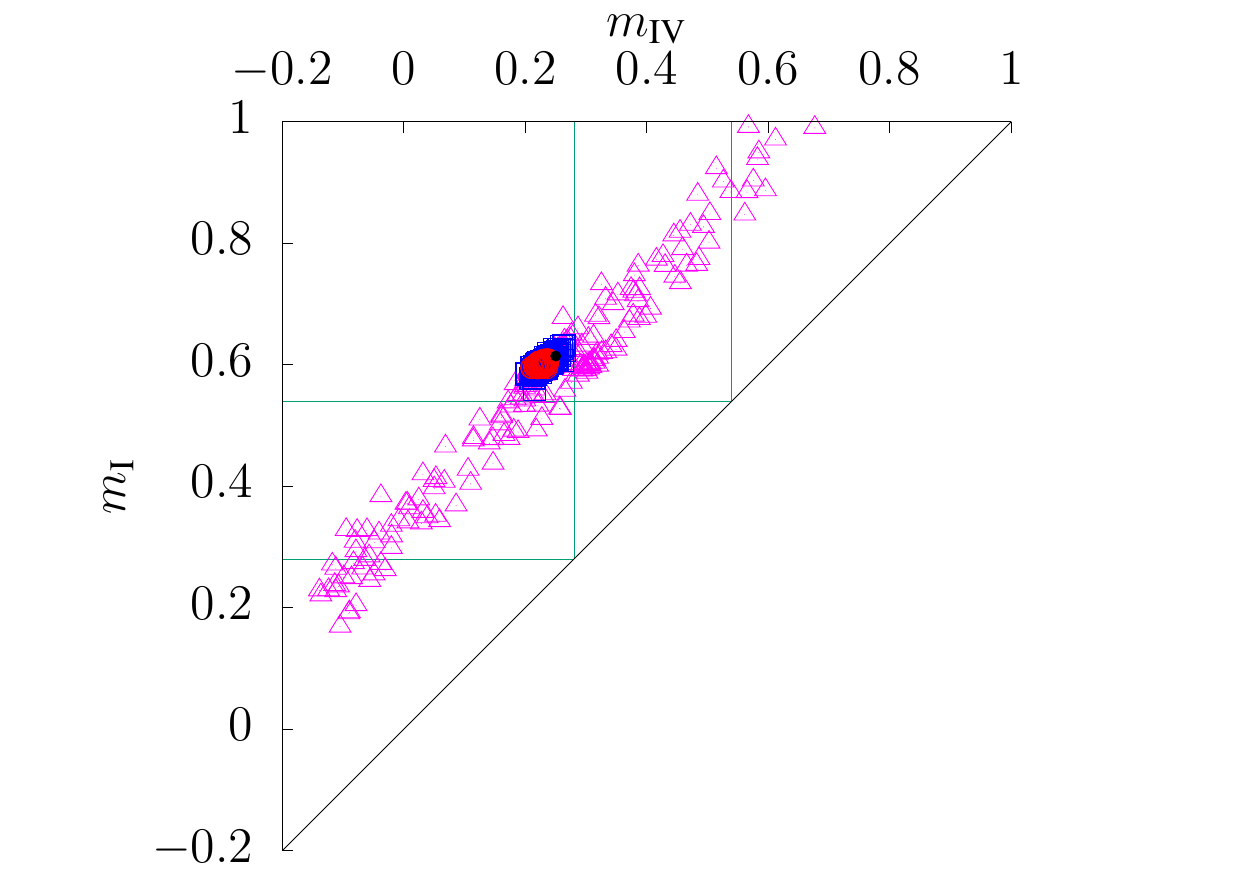}} \caption{} \label{fig:triangle_c} \end{subfigure}
    \caption{Schematic of the triangle theory (a); the ($m_\text{II},m_\text{III}$) plane is divided into different separation regions; $wx, wy, xy$ shows the boundary of the complete separation region. The neither pure raffinate nor pure extract regions is denoted as $\mathcal{E}$. Bayesian based triangle theory of the $(m_\text{II}, m_\text{III})$ plane (b) and the $(m_\text{IV}, m_\text{I})$ plane (c); the sample ensemble is illustrated with magenta triangles, while samples of $99\%$ purities with blue squares and samples of $99.9\%$ with red dots.}
    \label{fig:triangle}
\end{figure}

We are interested in determining the optimal operating conditions, $\theta$, of the four-zone \textsc{smb} process, for a given feed composition (\ie, $c_{\text{in},i}^F = \SI{550}{\gram\per\litre}\allowbreak = \SI{3.0528e3}{\mole\per\cubic\metre},\allowbreak\, i\in\{\text{glc}, \text{fru}\}$) in this study.
By using the triangle theory (provided the feed concentration is diluted such that nonlinear effects are negligible), the space of the operating parameters (\ie, the dimensionless flowrate ratios, $m_j,\, j\in\{\text{II}, \text{III}\}$) is divided into four major regions with different separation regimes (illustrated as $\mathcal{A}$, $\mathcal{B}$, $\mathcal{C}$, $\mathcal{D}$ in Fig.~\ref{fig:triangle_a}).
The boundaries (\ie, $wx, wy$) of the complete separation region, $\mathcal{A}$, are calculated by Eq.~\eqref{eq:triEq}, which involves the adsorption parameters of glucose and fructose, $H_{\text{glc}}$ and $H_{\text{fru}}$.
\begin{equation}\begin{split}
    H_{\text{glc}} &< m_{\text{II}} < m_{\text{III}} < H_{\text{fru}} \\
    m_j &= \frac{t_s Q^j - \varepsilon_t V_c}{(1 - \varepsilon_t) V_c}, \quad j \in\{\text{II}, \text{III}\}
    \label{eq:triEq}
\end{split}\end{equation}
where overall void fraction of the bed $\varepsilon_t = \varepsilon_c + \varepsilon_p(1 - \varepsilon_c)$, and $V_c$ denotes the volume of the chromatographic column.
By scanning the points in the complete separation region of the ($m_{\text{II}},m_{\text{III}}$) plane, the optimal conditions can be located. 

The complete separation region in the ($m_{\text{II}},m_{\text{III}}$) plane of the triangle theory was defined merely in terms of the purity performance indicator, that is, maximizing purities at the outlet streams with respect to the operating conditions $\theta$.
In this study, the samples drawn from the \textsc{mcmc} algorithm are defined with respect to the multi-objective function (cf.~Eq.~\eqref{eq:eps_constraint}).
All the collected samples (\ie, magenta triangle symbol in Fig.~\ref{fig:triangle_b}) from the multivariate posterior distribution are plotted onto the ($m_{\text{II}},m_{\text{III}}$) plane.
It is clearly shown that there is a strong linear correlation between the dimensionless ratios (the data is shown in \ref{app:corr}).
As $m_\text{II}$ and $m_\text{III}$ are linearly correlated, so no samples are located inside the $\mathcal{D}$ region.
In Fig.~\ref{fig:triangle_b}, the samples that render $99\%$ purities at both withdrawn outlets are illustrated with blue squares, while the $99.9\%$ purities are illustrated with red dots.
All these samples are located inside the complete separation region, $\mathcal{A}$, from the triangle theory.
The black dot (\ie, $(0.308, 0.484)$) on the $(m_\text{II}, m_\text{III})$ plane is the optimum point from \citet{klatt2002model}.
It not located inside the ellipse of $99.9\%$ purity samples, but in the ellipse of $99\%$ samples.
It is partially because of that the column length was optimized.
Moreover, these samples are far away from both the vertex ($w$), the diagonal ($xy$) and the boundaries ($wx, wy$) of the complete separation region.
This also implies robustness of the operating conditions, as explained in detail in \citet{mazzotti1997optimal} that the minimal distance of the operating point from the boundaries of the complete separation region can be a measure of the maximal acceptance perturbation on the values of operating conditions.
As seen from the sample density of different regions of the ($m_{\text{II}}, m_{\text{III}}$) plane, operating points are more easily to be located in the $\mathcal{C}$ region than that in the $\mathcal{B}$ region.
Therefore, from the \textsc{smb} simulation point of view, it is more possible to encounter chromatograms upon \textsc{css} with pure raffinate streams and polluted extract streams (cf.~the chromatogram of point $a$ in Fig.~\ref{fig:chromas}).

\begin{figure}
    \centering
    \includegraphics[width=0.49\textwidth]{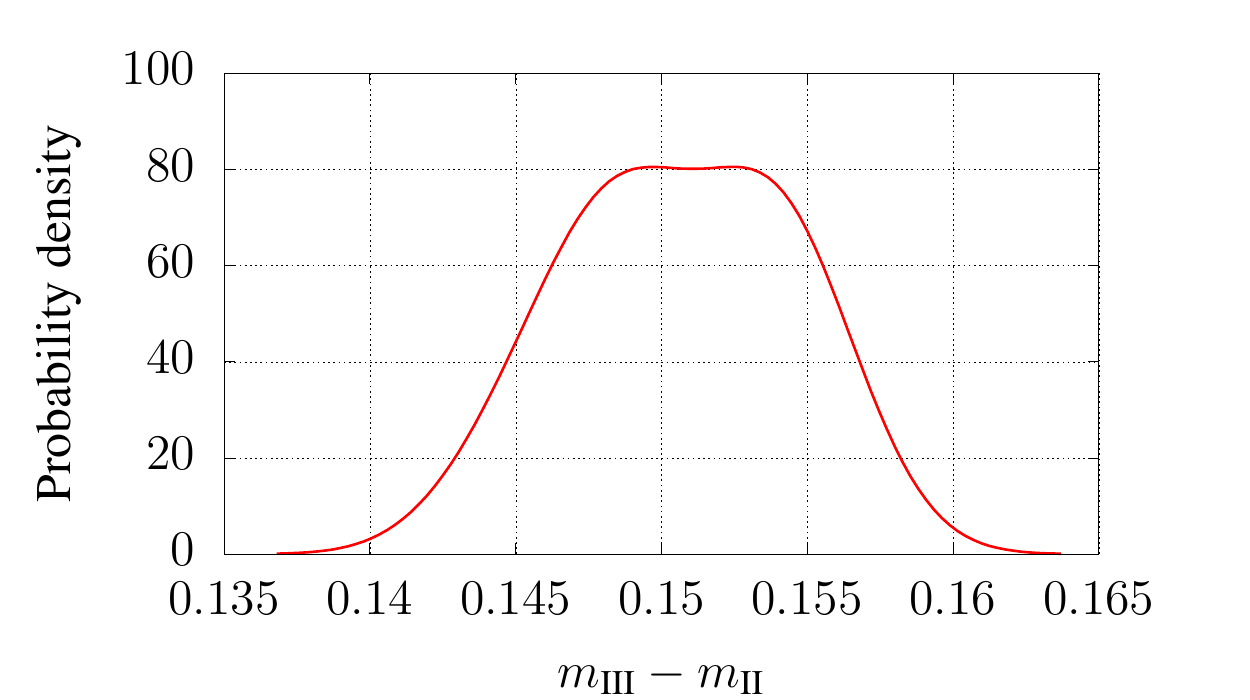}
    \includegraphics[width=0.49\textwidth]{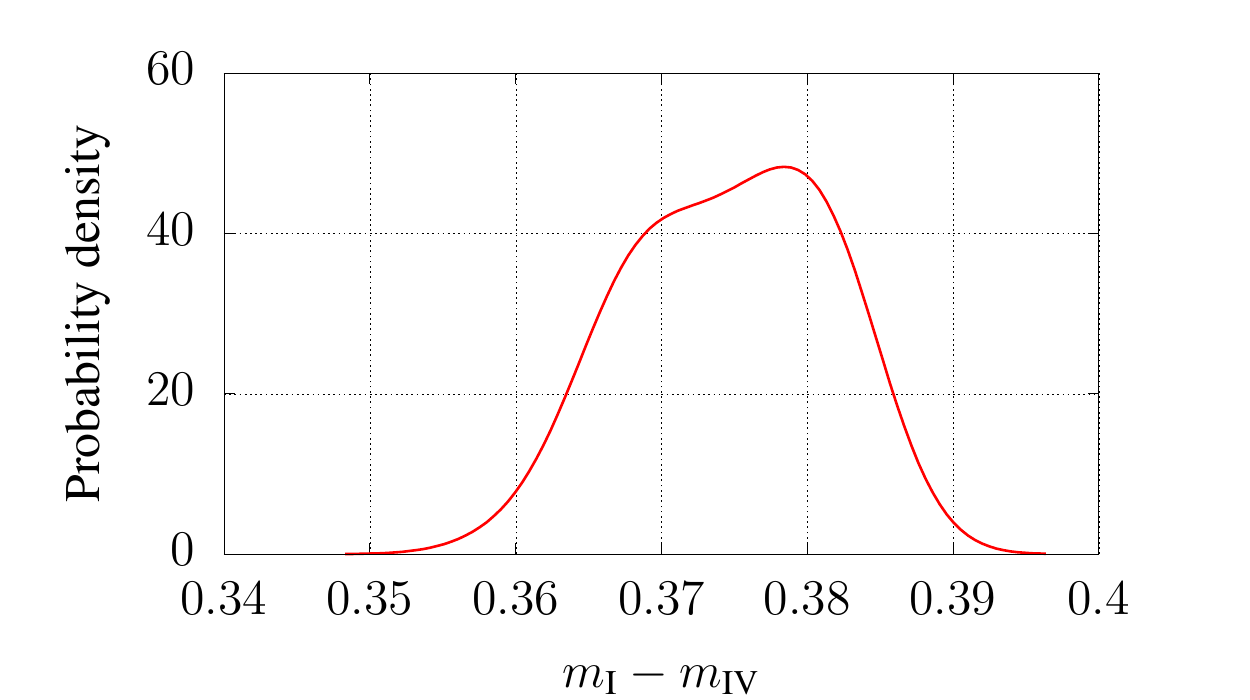}
    \caption{Probability distribution of the dimensionless ratio difference $m_\text{III} - m_\text{II}$ (left) and $m_\text{I} - m_\text{IV}$ (right).}
    \label{fig:m3m2}
\end{figure}

There is a cluster of optimal points merely in terms of the purity in the $\mathcal{A}$ region; they can be further distinguished according to other performance indicators.
$m_\text{III} - m_\text{II}$ denotes the amount of fresh feed that is treated in the \textsc{smb} process.
The purity performance improves by increasing the $m_\text{III} - m_\text{II}$ difference, that is, moving the operating points towards the vertex, $w$, across straight lines parallel to the diagonal in the ($m_\text{II}, m_\text{III}$) plane.
On the contrary, the yield performance enhances by decreasing the $m_\text{III} - m_\text{II}$ difference; the pressure drops also limit the difference.
As seen from Fig.~\ref{fig:m3m2}, the $m_\text{III} - m_\text{II}$ that results in optimal conditions follows Gaussian distribution with mean value of \num{0.15}.
It is located in the middle of the vertex, $w$, and the diagonal line, $xy$.
Furthermore, it was observed that more numerical iterations were required to converge to \textsc{css}, when the operating points are close to boundaries of each region.

It is worth considering the $(m_\text{II}, m_\text{III})$ plane of \textsc{smb} units, which plays a key role in performing the separation.
However, we also need to examine the projection of the four-dimensional region of the complete separation onto the $(m_\text{IV}, m_\text{I})$ plane.
A similar discussion applies to the $(m_\text{IV}, m_\text{I})$ semi-infinite plane (\ie, $m_\text{IV} < H_\text{glc}, H_\text{fru} < m_\text{I}$) (cf.~Fig.~\ref{fig:triangle_c}).
The black dot $(0.250, 0.614)$ is the optimum point from \citet{klatt2002model}.
Linear correlation is also observed, see \ref{app:corr}.
The distribution of $m_\text{I} - m_\text{IV}$ is shown in Fig.~\ref{fig:m3m2}; the value with the maximal probability is \num{0.378}.
As observed in Fig.~\ref{fig:triangle_c}, the distance from the ellipse centre to the $w$ vertex (\ie, 0.09) is shorter than that in the ($m_\text{III}, m_\text{II}$) plane (\ie, 0.13).
Thus, the ellipse is smaller and closer to the boundaries, in other words, less acceptance perturbation on the values of these operating conditions.
If we consider the acceptance ratio of the \textsc{mcmc} sampling (the samples located in $\mathcal{D}$ over all samples), there is higher rejection rate in the ($m_\text{IV}, m_\text{I}$) plane.
It is also worth mentioning that the short distance from the ellipse centre to the vertex is partially due to the performance of desorbent consumption is not taken into account in the objective function.

As have discussed in the previous subsection, advantages of using Bayesian inference in uncertainty analysis is inherent.
Advantages of Bayesian inference over the triangle theory in the determination of operating conditions is not obvious in this case study, where the linear isotherm is used and explicit algebraic expressions can be derived to calculate operating conditions. 
But, after having designed the complete separation region using the triangle theory, multiple operating points still need to be tested in order to pinpoint the optimal one in terms of performance indicators considered \citep{lubke2007numerical, heinonen2018chromatographic}.
Moreover, in nonlinear cases (\eg, steric mass-action kinetics), the potential of Bayesian inference is prominent, where the complete separation region is strongly asymmetrical, curvilinear and the triangle theory even can not be applied.
Nevertheless, the triangle theory is not trivial.
It is most useful for the investigation of separation carried out under linear and quasi-linear conditions.
Additionally, it provides a useful starting point for further nonlinear studies, as the condition of ideality and linearity can be relaxed.

\subsection{Multiple chain algorithm}

As have described in \emph{section}~\ref{sec:case}, the multiple chain \textsc{mcmc} algorithm was not run intrinsically; instead, two simulation instances were run simultaneously on the Linux node, with an additional program checking the convergence periodically.

An inherent multiple chain \textsc{mcmc} algorithm, differential evolution Markov Chain (\textsc{de-mc} \citep{braak2006markov, vrugt2009accdlerating}), has also been applied in this study.
\textsc{de-mc} is a combination of the differential evolution \citep{storn1997differential} with an added Metropolis step, in which multiple chains are run in parallel.
Thus, it can be effective to explore multi-modal densities.
20 multiple chains were applied.
The resulting posterior distributions of the operating conditions of the four-zone \textsc{smb} process convey the same information (\eg, Fig.~\ref{fig:paretos} and Fig.~\ref{fig:postDis}) as the ones from the presented \textsc{mcmc} algorithm (data is not shown).
As have illustrated in Fig.~\ref{fig:postDis}, all the distributions are unimodal; therefore, both the \textsc{mcmc} and the \textsc{de-mc} algorithms render the same results.
Nevertheless, slightly differences of the smoothness of posterior distributions are observed, which is attributed to the use of different mechanism of proposal distributions.
However, the performance and effectiveness of multiple chain \textsc{mcmc} algorithms could differ in process designs of \textsc{smb} models with nonlinear adsorption isotherms or kinetics.

\section{Conclusions}

Any type of uncertainties both in the batch and the \textsc{smb} chromatographic processes can affect their performance evaluation such that decision-making in application.
Discrepancies between experimental and simulation results have commonly observed in \textsc{smb} processes.
Moreover, \textsc{smb} industries have not been fully benefited from  mechanistic model-based methods.
These can often be attributed to the absence of uncertainty consideration.
To this end, a Bayesian inference framework for the uncertainty assessment in the process design of \textsc{smb} units has been introduced.
A classical glucose-fructose four-zone \textsc{smb} process under linear condition has been used as the example.
A Markov Chain Monte Carlo algorithm, that is, Metropolis algorithm incorporated with delayed rejection and adjusted Metropolis strategies, has been used for sampling.
The proposed method renders versatile information, \eg, well-determined operating conditions with credible intervals, robust posterior distributions, Pareto fronts and mapping to the triangle theory.

Although an \textsc{SMB} process with the adsorption behaviour described by the linear isotherm was used as an example, the power of proposed method is not limited to the linear situations.
It is more powerful in the nonlinear scenarios.
For instance, by using the Bayesian inference based methods, the separation regions of the triangle theory under strongly nonlinear conditions (where the triangle theory can not be applied) can be drawn. 
However, the computational cost, for cases with nonlinear kinetics and complex network configurations with a large number of columns, is expensive.
The \textsc{mcmc} algorithm applied in this study is not implemented and coded in an intrinsic parallel manner. 
Though having compared the proposed \textsc{mcmc} algorithm to an inherent parallel algorithm leads to no different results of, such as, posterior distributions and credible intervals, the potential of multi-chain \textsc{mcmc} algorithms should be more prominent in cases of multi-modal target density.

\section{Acknowledgements}
This work is supported by the National Key Research and Development Program of China under Grant No.~2019YFD0901805 and 2017YFB0309302.

\appendix

\section{Adaptive Metropolis} \label{app:covariance}

A random draw from the proposal distribution of Metropolis algorithm with an initially given covariance $\Sigma_0 = R_0 R_0^T$ is

\begin{equation}
    \tilde{\theta} = \theta + R_0 z,
    \label{eq:random_walk}
\end{equation}
where $z \in \mathbb{R}^n$ is an independent multivariate standard normal random vector.
An initial covariance matrix $\Sigma_0$ is approximated with regard to the initial point, $\theta_0$, by using the Fisher information matrix:

\begin{equation}
    \begin{split}
        \Sigma_0 &\approx \left( \left(\frac{\partial c}{\partial \theta}(z\!=\!L, t, \theta_0) \right)^T \Lambda^{-1} \frac{\partial c}{\partial \theta}(z\!=\!L,t, \theta_0) \right)^{-1} \\
        &= \tilde{\sigma}_0 \left( \left( U S V^T \right)^T \left( U S V^T \right) \right)^{-1} = \tilde{\sigma}_0 \left( V S^T U^T U S V^T \right)^{-1} \\
        &= \tilde{\sigma}_0 \left( V S^T S V^T \right)^{-1} \\
        &= \tilde{\sigma}_0\,  V \underbrace{ (S^T S)^{-1}}_{=[\text{diag}(S)]^{-2}} V^T
    \end{split}
    \label{eq:covar}
\end{equation}
Here, the weight matrix $\Lambda$ is $\text{diag}(\tilde{\sigma}_0)$ and the Jacobian matrix $\frac{\partial c}{\partial \theta}(z\!=\!L,t,\theta_0)$ is calculated with automatic differentiation \citep{puttmann2016utilizing}.
It was then decomposed into $U S V^T$ using a singular value decomposition.
The properties of $U^T U = I$ and $V^{-1} = V^T$ were used in the derivation of Eq.~\eqref{eq:covar}.

\section{Delayed rejection} \label{app:dr}

The acceptance probability of the second stage candidate was computed in a way such that the reversibility of the Markov chain is preserved.
Suppose the current position is $\theta^{i-1}$.
A new candidate $\theta^i$, which was generated by the proposal distribution $J_1(\theta^i|\theta^{i-1})$, is denied by the Metropolis probability.
Upon rejection, a second stage candidate $\theta^{i+1}$ is tested, which depends not only on the current position $\theta^{i-1}$ but also the denied position $\theta^{i}$.
The proposal is drawn from a different proposal distribution $J_2(\theta^{i+1}|\theta^{i-1}, \theta^i)$.
The chance to accept the second stage candidate is determined by $\beta(\theta^{i-1}, \theta^i, \theta^{i+1}) =\allowbreak \min\pbk{1,\, \gamma(\theta^{i-1}, \theta^i, \theta^{i+1})}$, where the ratio is

\begin{equation}
    \begin{split}
        & \gamma(\theta^{i-1}, \theta^i, \theta^{i+1}) \\
        & = \frac{p(\theta^{i+1}|\Psi) J_1(\theta^{i}|\theta^{i+1}) J_2(\theta^{i-1}|\theta^i, \theta^{i+1}) [1-\beta( \theta^i,\theta^{i+1})]}{p(\theta^{i-1}|\Psi) J_1(\theta^{i}|\theta^{i-1}) J_2(\theta^{i+1}|\theta^i, \theta^{i-1}) [1-\beta(\theta^i,\theta^{i-1})]} \\
        & = \frac{p(\theta^{i+1}|\Psi) J_1(\theta^{i}|\theta^{i+1}) [1-\beta(\theta^{i},\theta^{i+1})]}{p(\theta^{i-1}|\Psi) J_1(\theta^{i}|\theta^{i-1}) [1-\beta(\theta^{i},\theta^{i-1})]} \\
        & = q_1 \, q_2 \, \frac{1-\beta(\theta^{i},\theta^{i+1})}{1-\beta(\theta^{i},\theta^{i-1})},
    \end{split}
\end{equation}
where a symmetric proposal distribution (\ie $J_2$) was used as in the Metropolis algorithm.
Therefore, $J_2$ is independent of the rejected position, $\theta^i$, such that $J_2(\theta^{i-1}|\theta^i, \theta^{i+1}) = J_2(\theta^{i+1}|\theta^i, \theta^{i-1})$.
The ratios $q_1$ and $q_2$ are given by
\begin{equation}
    \begin{split}
    q_1 & = \frac{p(\theta^{i+1}|\Psi)}{p(\theta^{i-1}|\Psi)} \\
    & = \exp\cbk{ -\frac{1}{2} \pbk{ \mathcal{H}(\theta^{i+1}; d_k) - \mathcal{H} (\theta^{i-1}; d_k)} } \frac{p(\theta^{i+1})}{p(\theta^{i-1})}
    \end{split}\label{eq:DR_ratio_q1}
\end{equation}
and
\begin{equation}
    \begin{split}
    q_2 & = \frac{J_1(\theta^{i}|\theta^{i+1})}{J_1(\theta^{i}|\theta^{i-1})} \\
    &= \exp \left(- \frac{1}{2} \left[\left \| R^{-1} (\theta^{i+1}-\theta^{i}) \right\|^2 - \left \| R^{-1} (\theta^{i-1}-\theta^{i}) \right\|^2 \right] \right).
    \end{split}\label{eq:DR_ratio_q2}
\end{equation}
In Eq.~\eqref{eq:DR_ratio_q2}, we have assumed a multivariate normal proposal density $J_1$ with covariance matrix $\Sigma = RR^T$.

\section{Autocorrelation plot of MCMC samples}\label{app:autocorr}
\begin{figure}
    \centering
    \begin{subfigure}{0.49\textwidth}{\includegraphics[width=\textwidth]{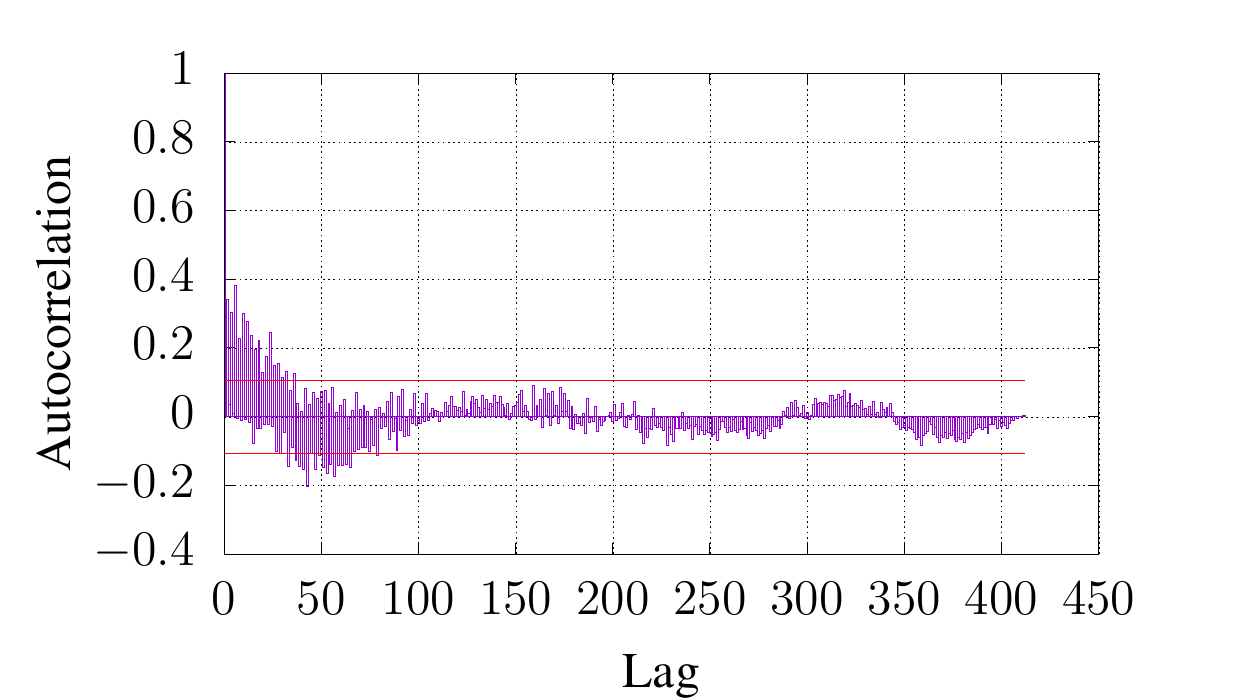}} \caption{$L$} \label{fig:autocorr_a} \end{subfigure}
    \begin{subfigure}{0.49\textwidth}{\includegraphics[width=\textwidth]{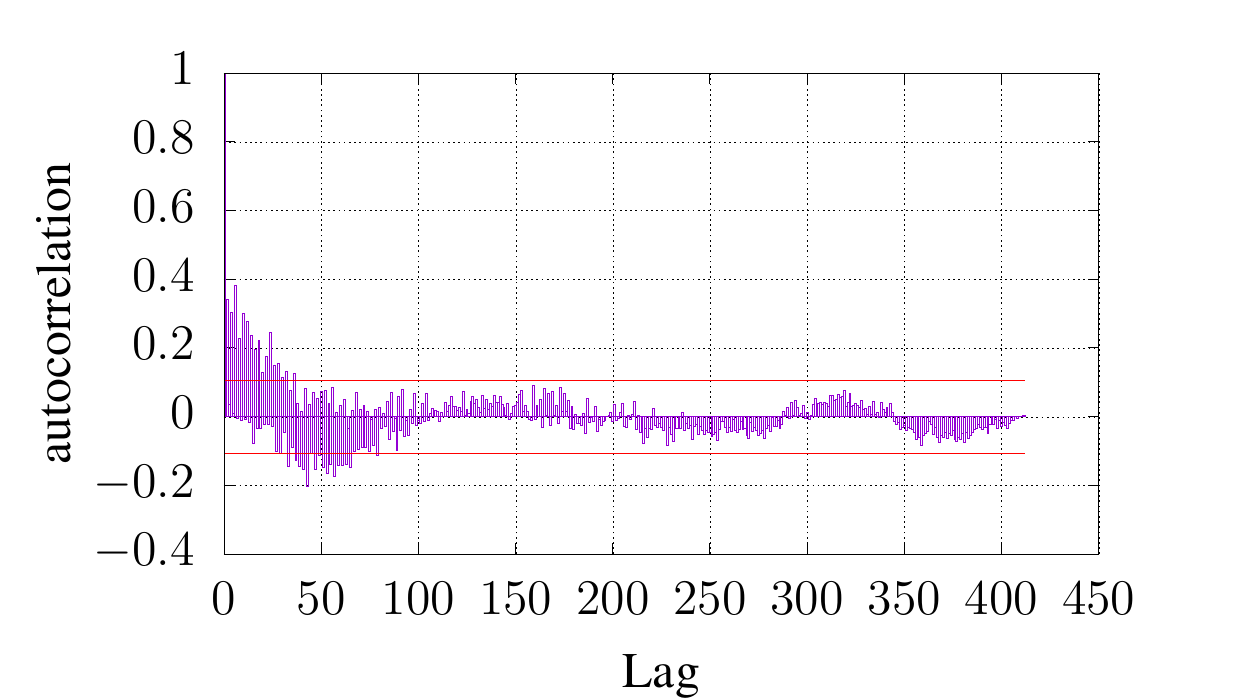}} \caption{$t_s$} \label{fig:autocorr_b} \end{subfigure}
    \begin{subfigure}{0.49\textwidth}{\includegraphics[width=\textwidth]{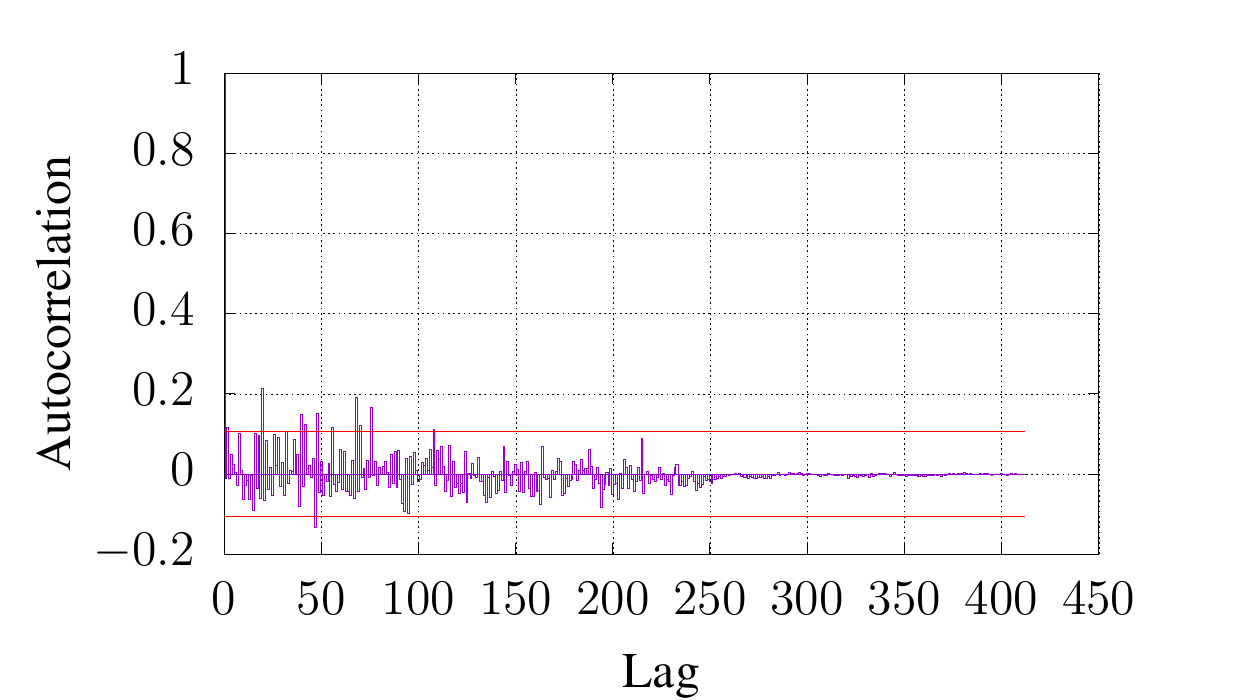}} \caption{$Q^\text{rec}$} \label{fig:autocorr_c} \end{subfigure}
    \begin{subfigure}{0.49\textwidth}{\includegraphics[width=\textwidth]{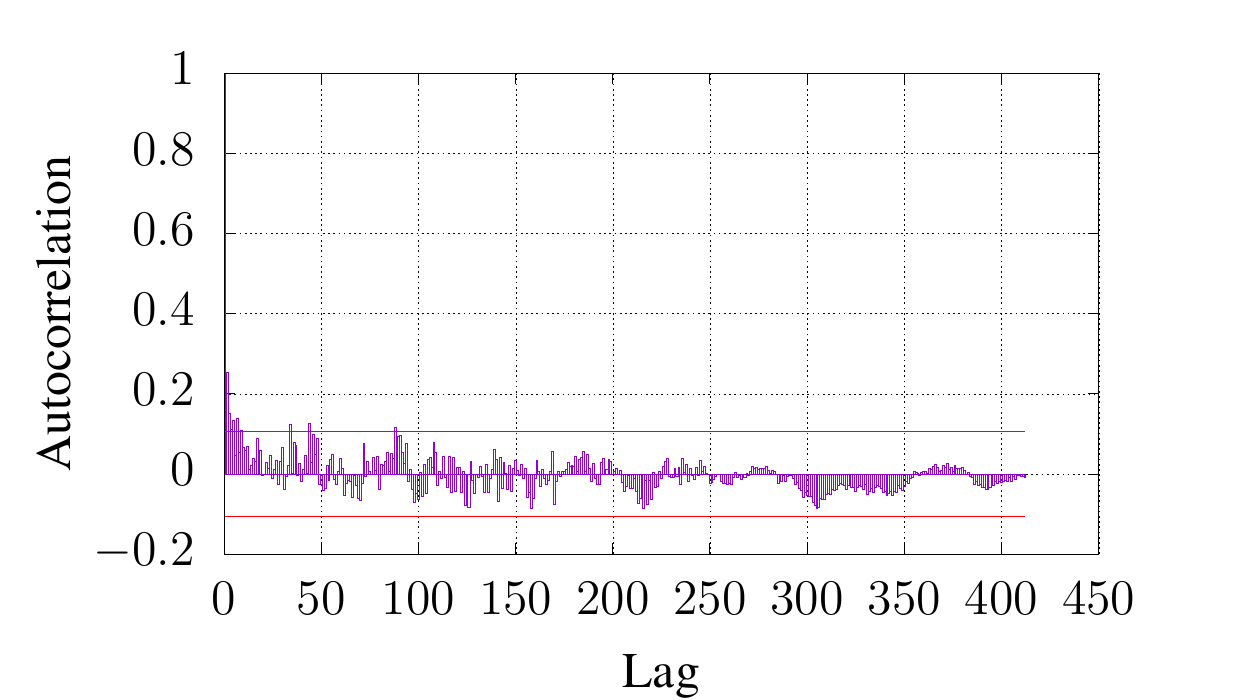}} \caption{$Q^F$} \label{fig:autocorr_d} \end{subfigure}
    \begin{subfigure}{0.49\textwidth}{\includegraphics[width=\textwidth]{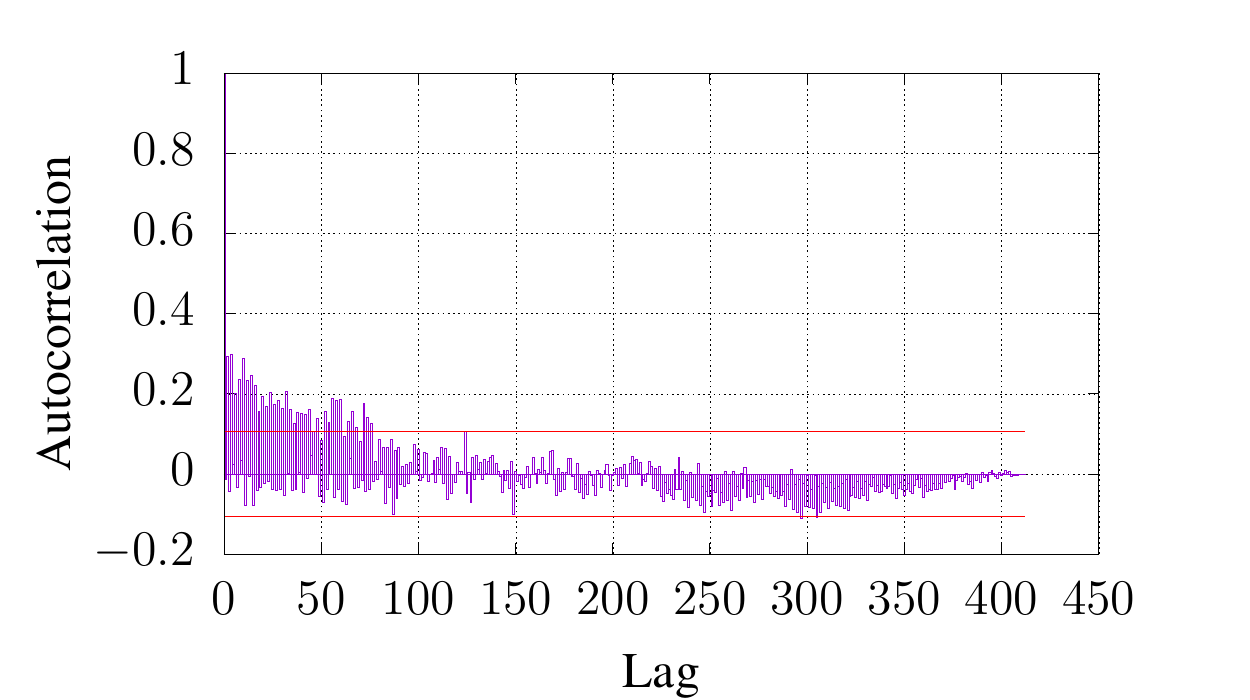}} \caption{$Q^D$} \label{fig:autocorr_e} \end{subfigure}
    \begin{subfigure}{0.49\textwidth}{\includegraphics[width=\textwidth]{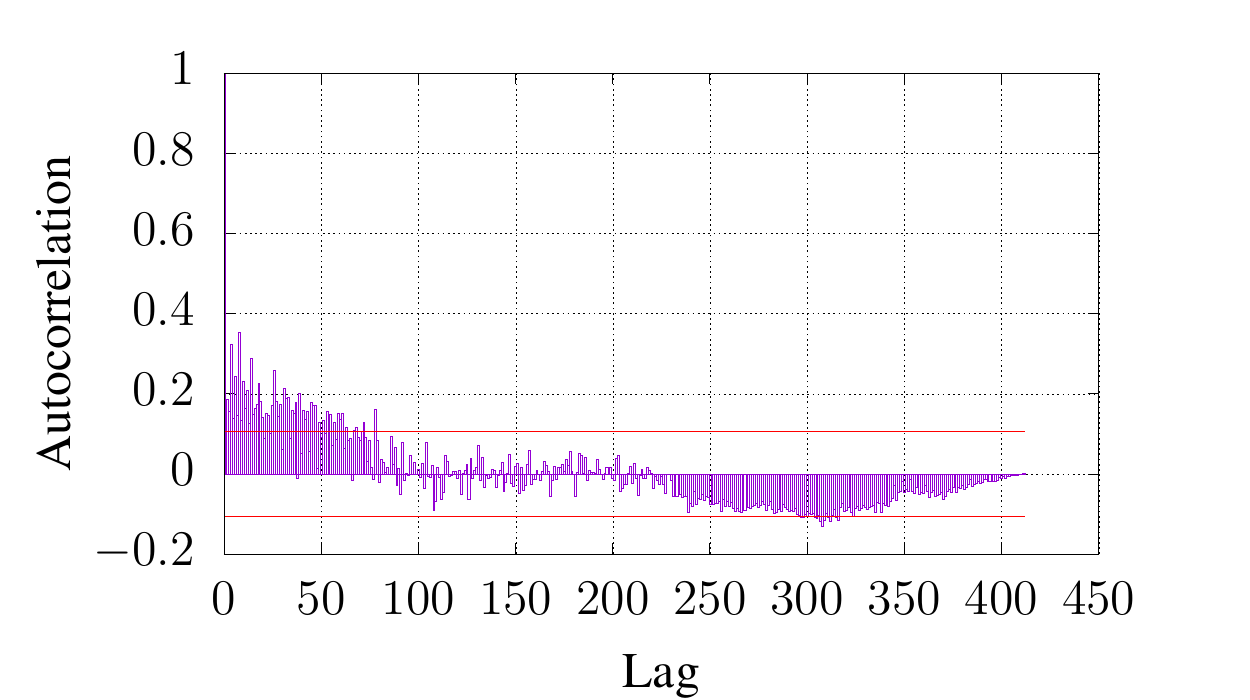}} \caption{$Q^E$} \label{fig:autocorr_f} \end{subfigure}
    \caption{Autocorrelation plot of the operating conditions.}
    \label{fig:autocorr}
\end{figure}

Autocorrelation, $\rho_t$, refers to the degree of correlation between the values of the same parameters across different observations (\ie, lag $t$).
The pattern of autocorrelation of each parameter, $\theta_\ell$, is illustrated in Fig.~\ref{fig:autocorr}.
As seen from Fig.~\ref{fig:autocorr}, when the lag $t>250$, the autocorrelation values $\rho_t$ get small enough.
It can be interpreted as that, upon convergence, the samples generated from the stationary distribution are random and not correlated with each other.

\section{Linear correlations of $m_j$}\label{app:corr}
\begin{figure}
    \centering
    \includegraphics[width=0.9\textwidth]{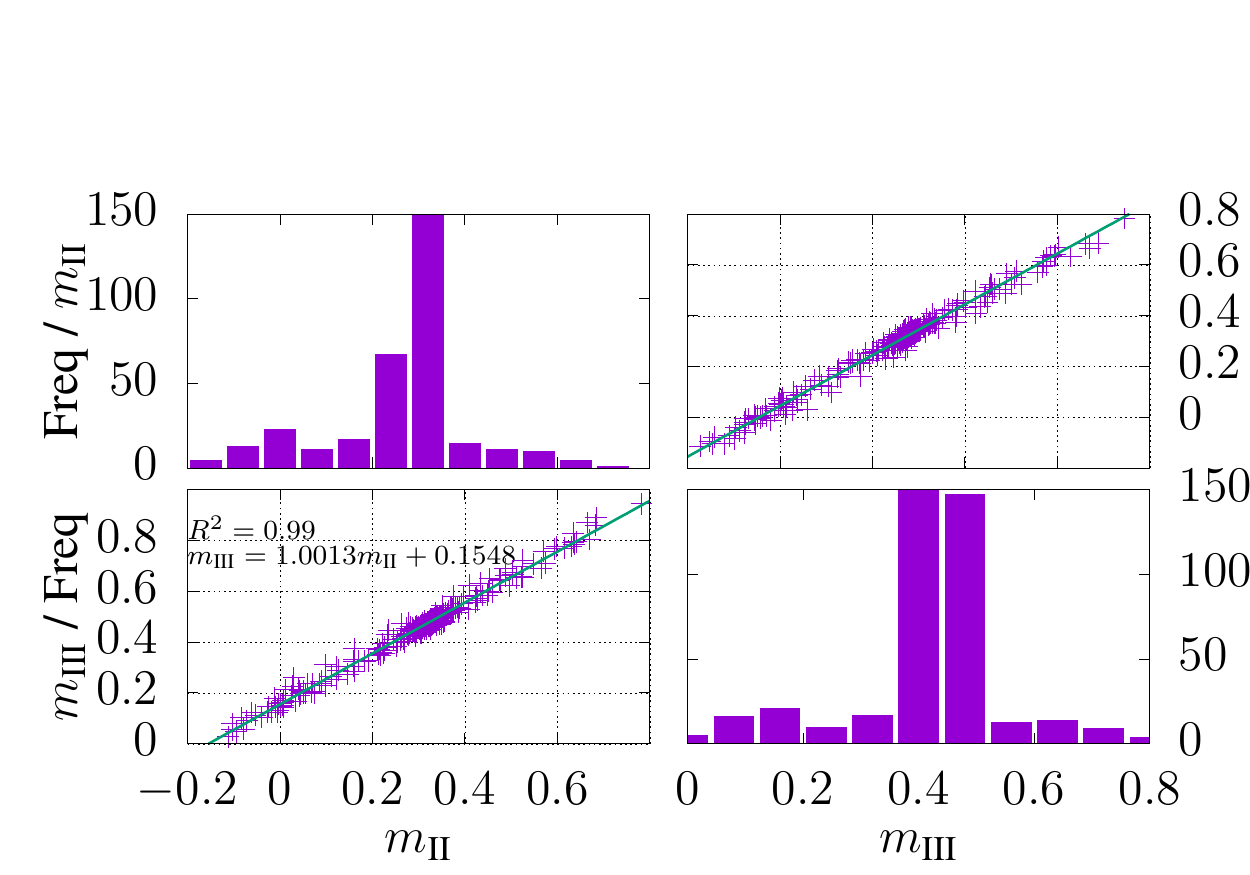}
    \includegraphics[width=0.9\textwidth]{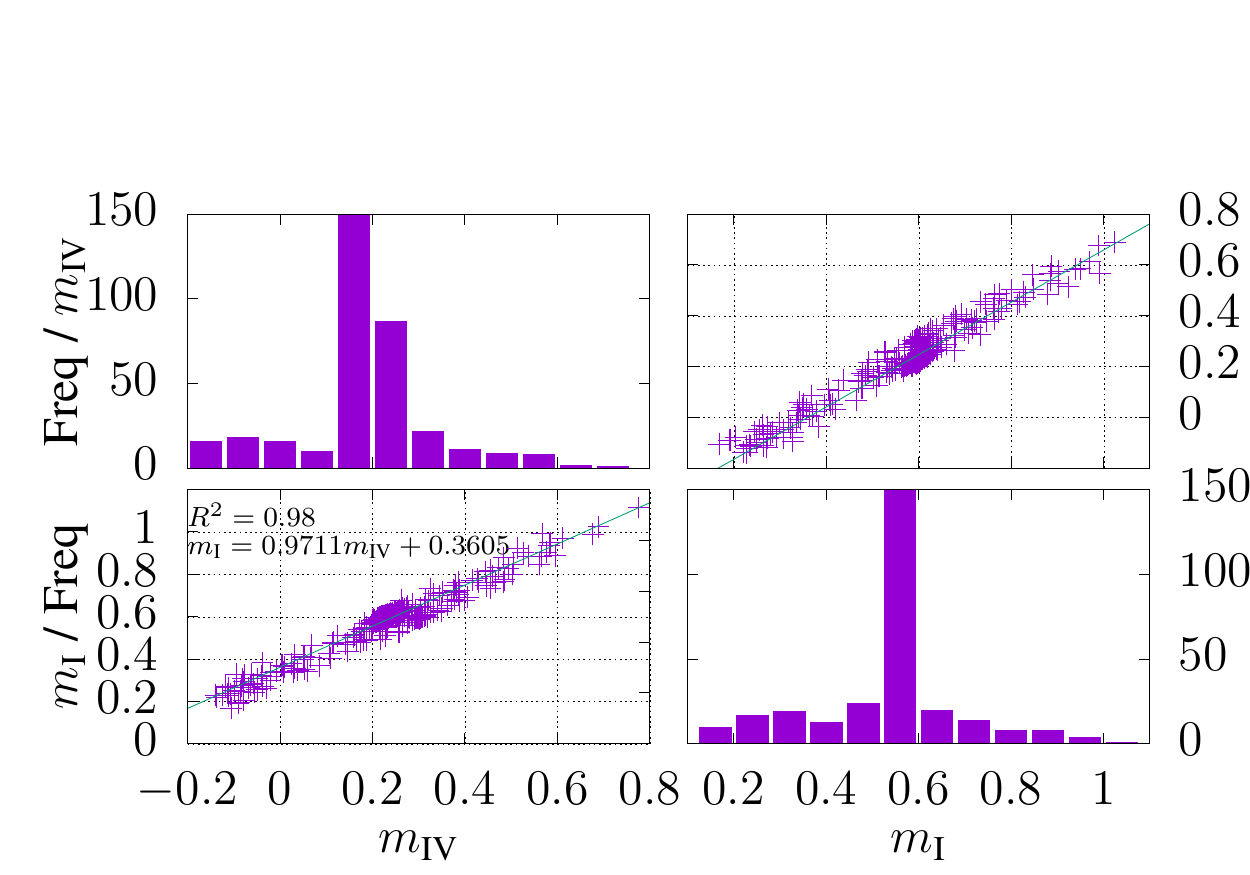}
    \caption{Linear correlation between the dimensionless values $m_\text{II}, m_\text{III}$ (top) and $m_\text{IV}, m_\text{I}$ (bottom).}
    \label{fig:corr}
\end{figure}

A linear correlation is observed between the dimensionless flowrate ratios, $m_\text{III}$ and $m_\text{II}$ (cf.~Fig.~\ref{fig:corr}).
The top left and bottom right figures show the histograms of $m_\text{II}$ and $m_\text{III}$, respectively.
The top right and bottom left figures both illustrate the linear correlation between $m_\text{III}$ and $m_\text{II}$.
To be specific, $m_\text{III} = 1.0013 m_\text{II} + 0.1548$ and the coefficient determination of the fitting is $R^2 = 0.99$.
In addition, the difference of $m_\text{III} - m_\text{II}$, which denotes the amount of fresh feed that is treated in the \textsc{smb} process, is around \num{0.1548}.
The most probable $m_\text{II}$ value is \num{0.33}; while the most probable $m_\text{III}$ value is \num{0.48}.
Therefore, the constraint $H_\text{fru} > m_\text{III} > m_\text{II} > H_\text{glc}$ holds.

$m_\text{I}$ is also linearly correlated with $m_\text{IV}$, $m_\text{I} = 0.9711 m_\text{IV} + 0.3605$, with the coefficient determination $R^2 = 0.98$.
The difference of $m_\text{I} - m_\text{IV}$ denotes the amount of fresh desorbent that is treated in the \textsc{smb} process.
The most probable $m_\text{I}$ value is \num{0.60}, while the most probable $m_\text{IV}$ value is \num{0.23}.
Thus, the constraints $H_\text{fru} < m_\text{I}$ and $m_\text{IV} < H_\text{glc}$ are satisfied.

\bibliographystyle{elsarticle-num-names.bst}

\bibliography{references}

\end{document}

%% file: preamble.tex
\usepackage[utf8]{inputenc}

\usepackage{amsmath,amssymb,amsthm,amscd}
\usepackage{mathtools}
\usepackage{dsfont} 
\usepackage{bm} 
\usepackage[binary-units=true]{siunitx}
\usepackage{multirow}
\usepackage{caption}
\usepackage[font=footnotesize]{subcaption}
\usepackage{booktabs} 
\usepackage{blkarray} 
\usepackage{pdflscape}
\usepackage{longtable}

\usepackage{tikz}
\usepackage{xcolor}
\usepackage{graphicx} 
\usepackage[colorlinks=true, citecolor=blue, linkcolor=blue]{hyperref}

\usepackage{paralist} 
\usepackage{verbatim} 

\usepackage{fancybox}
\usepackage{varioref} 
\usepackage{chapterbib} 
\usepackage{eurosym}
\usepackage[running]{lineno}
\usepackage{afterpage}

\usepackage[compatible]{nomencl}
\usepackage{xspace}
\newcommand{\dd}{\mathrm{d}} 
\newcommand{\abs}[1]{\left\lvert#1\right\rvert}
\newcommand{\norm}[1]{\left\lVert#1\right\rVert}

\newcommand{\pbk}[1]{\left(#1\right)}
\newcommand{\cbk}[1]{\left\lbrace#1\right\rbrace}

\newcommand{\ie}{\textit{i.e.}\@\xspace}
\newcommand{\eg}{\textit{e.g.}\@\xspace}
\newcommand{\HE}{\hbox{H\kern-.12em\lower.48ex\hbox{E}}}
\def\leaderfill{\leaders\hbox to 1em{\hss.\hss}\hfill}
\DeclareSIUnit\molar{\mole\per\cubic\metre}
\DeclareSIUnit\Molar{\textsc{M}}